\documentclass[preprint,review,12pt]{elsarticle}
\makeatletter
\def\ps@pprintTitle{%
 \let\@oddhead\@empty
 \let\@evenhead\@empty
 \def\@oddfoot{}%
 \let\@evenfoot\@oddfoot}

\usepackage{amssymb}

\usepackage{xcolor}
\definecolor{blue}{RGB}{0,0,0}
\definecolor{orange}{RGB}{0,0,0}
\colorlet{brown}{black}

\usepackage{verbatim}
\usepackage{tabularx}
\usepackage[T1]{fontenc}
\usepackage{graphicx}
\usepackage{booktabs}
\usepackage{multirow}
\usepackage{hyperref}
\usepackage{rotating}

\usepackage[inline]{enumitem}
\usepackage{tabularx}
\usepackage{color}
\usepackage{todonotes}
\usepackage{tablefootnote}
\usepackage{longtable}
\usepackage{xspace}
\usepackage{makecell}
\usepackage{multirow}
\usepackage{units}
\usepackage{xcolor}
\usepackage{footnote}
\usepackage{colortbl}

\usepackage{pifont}
\newcommand{\xmark}{\ding{55}}%

\newcommand{\mc}{\textsc{mc}\xspace}
\newcommand{\iknn}{\textsc{iknn}\xspace}
\newcommand{\sknn}{\textsc{sknn}\xspace}

\newcommand{\smf}{\textsc{smf}\xspace}
\newcommand{\fpmc}{\textsc{fpmc}\xspace}
\newcommand{\bpr}{\textsc{bpr-mf}\xspace}
\newcommand{\fism}{\textsc{fism}\xspace}
\newcommand{\fossil}{\textsc{fossil}\xspace}

\newcommand{\ppop}{\textsc{ppop}\xspace}
\newcommand{\pop}{\textsc{pop}\xspace}
\newcommand{\mr}{\textsc{mr}\xspace}
\newcommand{\hrm}{\textsc{hrm}\xspace}
\newcommand{\swiwo}{\textsc{swiwo}\xspace}
\newcommand{\caser}{\textsc{caser}\xspace}
\newcommand{\samr}{\textsc{samr}\xspace}
\newcommand{\bprgrutwo}{\textsc{bpr-gru4rec2}\xspace}

\newcommand{\sr}{\textsc{sr}\xspace}
\newcommand{\usr}{\textsc{sr\_b}\xspace}

\newcommand{\usrreminder}{\textsc{sr\_br}\xspace}

\newcommand{\vsknn}{\textsc{vsknn}\xspace}
\newcommand{\uvsknn}{\textsc{vsknn\_eb}\xspace}
\newcommand{\vsknnreminder}{\textsc{vsknn\_r}\xspace}
\newcommand{\uvsknnreminder}{\textsc{vsknn\_ebr}\xspace}

\newcommand{\stan}{\textsc{stan}\xspace}
\newcommand{\staner}{\textsc{stan\_er}\xspace}
\newcommand{\stanebr}{\textsc{stan\_ebr}\xspace}
\newcommand{\stanr}{\textsc{stan\_r}\xspace}
\newcommand{\vstan}{\textsc{vstan}\xspace}
\newcommand{\vstanebr}{\textsc{vstan\_ebr}\xspace}
\newcommand{\vstaneb}{\textsc{vstan\_eb}\xspace}
\newcommand{\vstanr}{\textsc{vstan\_r}\xspace}

\newcommand{\gru}{\textsc{gru4rec}\xspace}
\newcommand{\grutwo}{\textsc{gru4rec2}\xspace}
\newcommand{\grureminder}{\textsc{gru4rec\_r}\xspace}

\newcommand{\narm}{\textsc{narm}\xspace}
\newcommand{\narmreminder}{\textsc{narm\_r}\xspace}

\newcommand{\hgru}{\textsc{hgru4rec}\xspace}
\newcommand{\iirnn}{\textsc{iirnn}\xspace}
\newcommand{\ncsf}{\textsc{ncsf}\xspace}
\newcommand{\nsar}{\textsc{nsar}\xspace}
\newcommand{\shan}{\textsc{shan}\xspace}

\newcommand{\reminders}{{reminders}\xspace}
\newcommand{\remindstrategy}{{remind\_strategy}\xspace}
\newcommand{\remindsess}{{remind\_sessions\_num}\xspace}
\newcommand{\wbase}{{weight\_Rel}\xspace}
\newcommand{\wirec}{{weight\_IRec}\xspace}
\newcommand{\wssim}{{weight\_SSim}\xspace}

\newcommand{\xing}{XING\xspace}
\newcommand{\retailrocket}{RETAIL\xspace}
\newcommand{\lastfm}{LASTFM\xspace}

\newcommand{\cosmetics}{COSMETICS\xspace}

\newcommand{\best}[1]{\textbf{#1}}
\newcommand{\sign}[1]{$*$\textbf{#1}}

\newcommand{\onesign}[1]{$*${#1}}
\newcommand{\scnd}[1]{\underline{#1}}

\journal{Information Sciences}

\begin{document}

\begin{frontmatter}



\title{Session-aware Recommendation: A Surprising Quest for the State-of-the-art}


\author[kl]{Sara Latifi}
\ead{sara.latifi@aau.at}
\address[kl]{University of Klagenfurt, Austria}
\author[to]{Noemi Mauro}
\ead{noemi.mauro@unito.it}
\address[to]{University of Torino, Italy}
\author[kl]{Dietmar Jannach}
\ead{dietmar.Jannach@aau.at}

\begin{abstract}
Recommender systems are designed to help users in situations of information overload. In recent years we observed increased interest in session-based recommendation scenarios, where the problem is to make item suggestions to users based only on interactions observed in an ongoing session, e.g., on an e-commerce site. However, in cases where interactions from previous user sessions are also available, the recommendations can be personalized according to the users' long-term preferences, a process called \emph{session-aware} recommendation. Today, research in this area is scattered, and many works only compare a newly proposed \emph{session-aware} with existing \emph{session-based} models. This makes it challenging to understand what represents the state-of-the-art.
To close this research gap, we benchmarked recent session-aware algorithms against each other and against a number of session-based recommendation algorithms along with heuristic extensions thereof. Our comparison, to some surprise, revealed that
\begin{enumerate*}[label=\textit{(\roman*)}]
\item simple techniques based on nearest neighbors consistently outperform recent neural techniques and that
\item session-aware models were mostly not better than approaches that do \emph{not} use long-term preference information.
\end{enumerate*}
Our work therefore points to potential methodological issues where new methods are compared to weak baselines, and it also indicates that there remains a huge potential for more sophisticated session-aware recommendation algorithms.

\end{abstract}


\end{frontmatter}

\section{Introduction}
Recommender systems (RS) can nowadays be found on many modern e-commerce and media streaming sites, where they 
help users find items of interest in situations of information overload. One reason for the success of RS lies in their ability to personalize the item suggestions based on the preferences and observed past behavior of the individual users. Historically, researchers have therefore strongly focused on situations where only information about long-term user preferences is available, e.g., in the form of item ratings. Only in recent years, more focus was put on the problem of \emph{session-based} recommendation, where the system has to deal with anonymous users and therefore can base its recommendations only on a small number of interactions that are observed in the ongoing session. 

Due to the practical relevance of this problem, a variety of technical approaches to session-based recommendation were proposed in the past few years, in particular ones based on deep learning (neural) techniques, see \cite{ludewigetal2019,wang2020survey}. 
Implicitly, these methods try to make recommendations by guessing the user's short-term intent or situational context only from the currently observed interactions.
However, while it is well known that the current intents and context may strongly determine which items are relevant in the given situation \citep{Jannachetal2015}, 
information about long-term preferences of users, if available, should not be ignored. In particular, the consideration of such information allows us to make session-based recommendations that are \emph{personalized} according to long-term preferences, a process which is also called \emph{session-aware} recommendation \citep{QuadranaetalCSUR2018}.

Session-aware recommendation problems are recently receiving increased interest. Today, the research literature is however still scattered, which makes it difficult to understand what represents the state-of-the-art in this area. One particular problem in that context is that existing works do not use a consistent set of baseline algorithms in their performance comparisons. Some works, for example, mainly compare session-\emph{aware} models with session-\emph{based} ones, i.e., with algorithms that do not consider long-term preference information, e.g., \gru \citep{Hidasi2016GRU}. Several other works use \emph{sequential} recommendation algorithms, e.g., \fossil \citep{Fossil2016},  as a baseline. These are algorithms that consider the sequence of the events but are usually designed for settings where the input is a time-stamped user-item rating matrix and not a sequential log of observed interactions. Only in a few works, previous session-aware algorithms are considered in the evaluations. One example is the method by Phuong et al.~\cite{NSAR2019}, which uses the \hgru method \citep{quadranahgru2017} as a baseline. Finally, almost all works include some trivial baselines, e.g., the recommendation of popular items in a session.

With this work, our goal is to close the research gap regarding the state-of-the-art in session-aware recommendation. For this purpose, we have conducted extensive experimental evaluations in which we compared five recent neural models for session-\emph{aware} recommendation with
\begin{enumerate*}[label=\textit{(\roman*)}]
\item a set of existing neural and non-neural approaches to session-\emph{based} recommendation, and
\item heuristic extensions of the session-based techniques that, e.g., make use of reminders \citep{JannachLudewigLerche2017umuai} or consider interactions that were observed immediately before the ongoing session.
\end{enumerate*}

Regarding the baseline techniques, we in particular considered methods based on nearest neighbors techniques, which previously proved to be very competitive in session-based recommendation scenarios \citep{ludewiglatifiumuai2020}. All investigated techniques were compared by extending the evaluation framework shared in \citep{Ludewig2018}. For reproducibility purposes, we share all data and code used in the experiments online\footnote{\textcolor{blue}{\url{ https://github.com/rn5l/session-rec/}}}, including the code for data preprocessing, hyperparameter optimization, and measuring.

The results of our investigations are more than surprising. In the majority of cases, and on all four considered datasets, heuristic extensions of existing \emph{session-based} algorithms were the best-performing techniques. In many cases, even plain session-based techniques, and in particular ones based on nearest-neighbor techniques, outperform recent session-aware models even though they do not consider the available long-term preference information for personalization. With our work, we therefore provide new baselines that should be considered in future works on session-aware recommendation.
On a more general level, our observations also point to potential methodological issues, where new models are compared to baselines that are either not properly optimized, that do not leverage all available information, or that are rather weak for the given task. Similar observations were previously made in the field of information retrieval and in other areas of recommender systems \citep{Armstrong:2009:IDA:1645953.1646031,ferraridacrema2020tois,Yang:2019:CEH:3331184.3331340}.

On a more positive note, our evaluations suggest that there is a huge potential to be tapped by more sophisticated (neural) algorithms that combine short-term and long-term preference signals for session-aware recommendation. An important prerequisite for progress in this area however lies in an increased level of reproducibility of published research. A side observation of our research is that despite some positive developments in recent years, where researchers increasingly share their code on public repositories, it in many cases still remains challenging to reproduce existing works.

The paper is organized as follows. In Section \ref{sec:related-work}, we discuss relevant previous works. 
Section \ref{sec:experiment-setup} describes our research methodology in more detail with respect to the compared algorithms, the evaluation protocol, and the performance measures. The results of our experiments are reported in Section \ref{sec:results}, \textcolor{blue}{and we discuss research implications and future works in Section \ref{sec:implications}}.

\section{Previous Work}
\label{sec:related-work}
Historically, recommender systems research focused strongly on the problem of rating prediction given a user-item rating matrix, a setting which is also known as the ``matrix completion'' problem \citep{Resnick:1994:GOA:192844.192905}. In this original collaborative filtering problem setting, the order of the ratings or the time when they were provided were not considered in the algorithms. Soon, however, it turned out that these aspects can matter, leading to the development of \emph{time-aware} recommender systems \citep{campos2014tars}, e.g., in the form of the \emph{timeSVD++} algorithm as used in the Netflix Prize \citep{Koren:09}.

Ten years after the Netflix Prize, the focus of research has mostly shifted from rating prediction to settings where mainly implicit feedback signals by users (e.g., purchase or item-view events) are available. Moreover, instead of considering the user-item matrix as the main input, recent research more often focuses on settings where the main input to a recommendation algorithm are sequential logs of recorded user interactions. The family of approaches that relies on such types of inputs are referred to as \textit{sequence-aware} recommender systems \citep{QuadranaetalCSUR2018}.

Within this class of sequence-aware recommender systems, we can differentiate between three main categories of problem settings and algorithmic approaches.

\begin{itemize}
    \item \textit{Sequential Recommender Systems:} Unlike the other types of sequence-aware approaches discussed here, these systems are rooted in the tradition of relying on a user-item rating matrix as an input. The particularity of such systems is that the sequence of events is first extracted from the time-stamped rating matrix, and the goal then usually is to predict the immediate next user action (e.g., the next Point-of-Interest that a user will visit), given the entire preference profile of the user.
    \item \textit{Session-based Recommender Systems:} The input to these systems are time-ordered logs of recorded user interactions, where the interactions are grouped into \emph{anonymous sessions}. Such a session could, for example, correspond to a listening session on a music service, or a shopping session on an e-commerce site. One particularity of such approaches is that users are not tracked across sessions, which is a common problem on websites that deal with first-time users or users that are not logged in. The prediction task in this setting is to predict the next user action, given only the interactions of the current session. Today, session-based recommendation is a highly active research area due to its practical relevance.
    \item \textit{Session-aware Recommender Systems:} This category is also referred to as \emph{personalized session-based recommender systems}. It is similar to session-based recommendation in that the user actions are grouped into sessions. Also the prediction goal is identical. However, in this problem setting, users are not anonymous, i.e., it is possible to leverage information about past user sessions when predicting the next interaction for the current session.
\end{itemize}
\begin{figure}[t]
\centering
 \includegraphics[width=.8\textwidth]{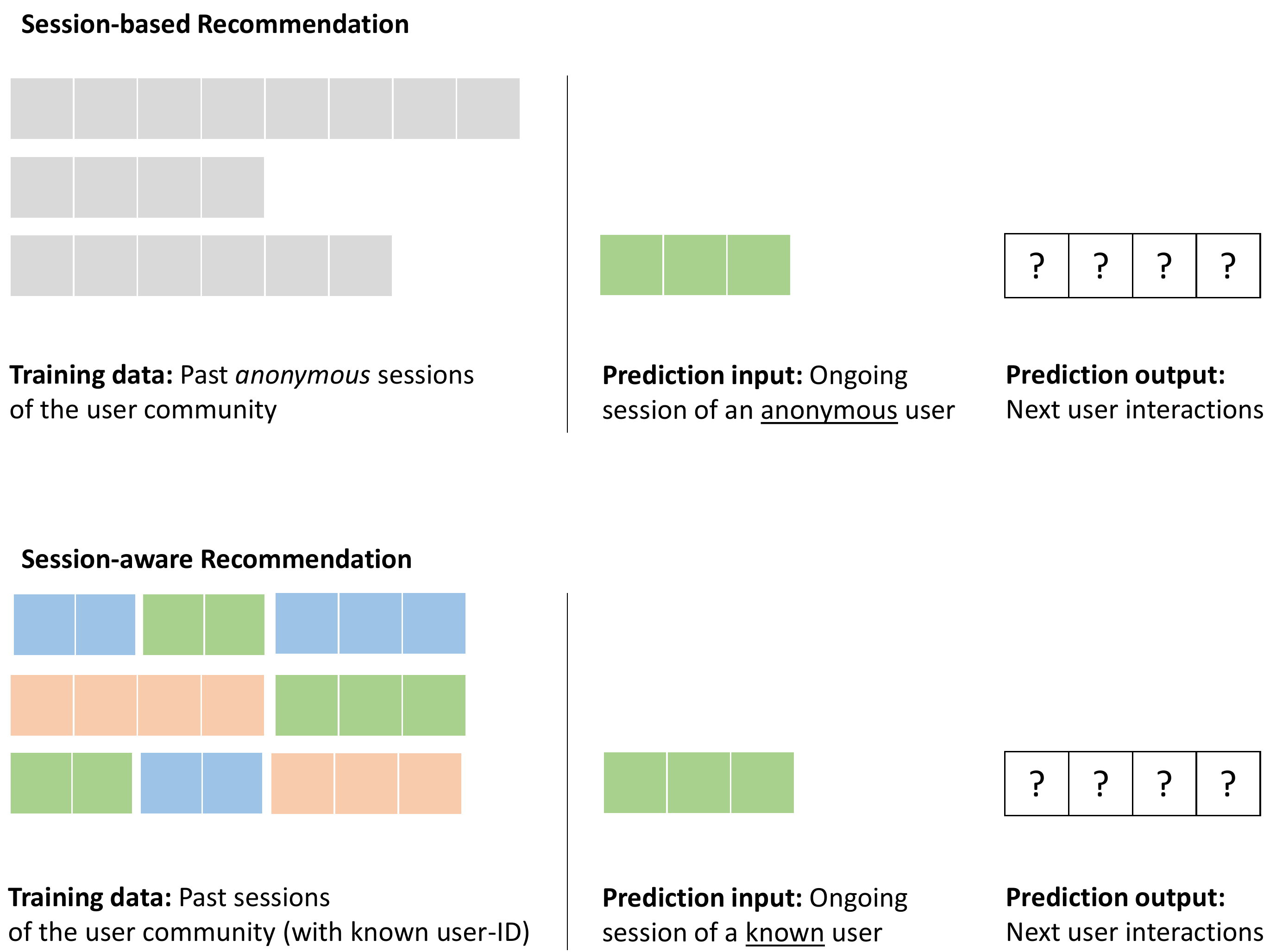}
 \caption{\textcolor{blue}{Comparison of session-based and session-aware recommendation problems. Different colors indicate different users.}}
 \label{fig:session-based-vs-session-aware}
\end{figure}

Figure \ref{fig:session-based-vs-session-aware} illustrates the differences between session-\emph{based} and session-\emph{aware} recommendation scenarios.\footnote{Remember that sequential problems, which are not based on the concept of a session, are not in the scope of this work.} In both problem settings, the recommendation problem consists of predicting which action a user will do next in an ongoing session. 
In an e-commerce setting, for example, the problem is to predict which items are relevant for the user, given the last few observed interactions in an ongoing shopping session.
The difference between the two scenarios however is that in one case we have long-term preference information about the user (session-aware recommendation), whereas in the other case such information is not available. Technically, this usually means that in session-aware scenarios we have user identifiers attached to the past usage sessions in the training data. In Figure \ref{fig:session-based-vs-session-aware}, we indicate sessions by different users with different colors.

Unfortunately, the terminology in the literature is not entirely consistent. In this work, we will therefore use the categorization and terminology as described above to avoid confusion. Next, we review the main technical approaches in each category.

\paragraph{Sequential Recommendation Approaches}

The first comparably simple approaches in this category were based on Markov models, e.g., \citep{markovchainnorris97}. Later on, more sophisticated approaches emerged which, for example, combined the advantages of matrix factorization techniques with sequence modeling approaches. Rendle et al.~\cite{Rendle2010} proposed the Factorized Personalized Markov Chain (\fpmc) approach as an early method for next-basket recommendation in e-commerce settings, where user interactions are represented as a three dimensional tensor (user, current item, next-item). Kabbur et al.~\cite{kabbur13fism} later proposed \fism, a method based on an item-item factorization. \fism was then combined with factorized Markov chains to incorporate sequential information into the \fossil model \citep{he16fusing}.

In recent years, several sequential recommender systems based on neural networks were developed. Kang and McAuley \cite{Kang:18}, for example, proposed \textsc{sasrec} (self-attention based sequential model), a method that allows to capture long-term semantics like an RNN. However, through the use of an attention mechanism, it focuses only on a smaller set of interactions to make the item predictions. In the \caser method, Tang and Wang \cite{Tang:18} embedded a sequence of recent items into latent spaces as an ``image'' in time, and proposed to learn sequential patterns as local features of the image with the help of convolutional filters. Most recently, Sun et al.~\cite{Sun:19} proposed \textsc{bert4rec}, which employs a deep bidirectional self-attention mechanism to model user behavior sequences.

In this work, we do not consider this class of algorithms in our performance comparison because these methods, in their original designs, do not consider the concept of a session in the input data. While it is in principle possible to apply these methods in a particular way for session-based recommendation problems, a previous evaluation shows that sequential approaches are often not competitive with techniques that were specifically designed for the problem setting. Specifically, the evaluation presented in \citep{Ludewig2018} included a number of sequential approaches, namely \fpmc, \mc, \smf, \bpr, \fism, and \fossil.\footnote{\fpmc: Factorized Personalized Markov Chains \citep{Rendle2010}, \mc: Markov Chains \citep{markovchainnorris97}, \smf: Session-based Matrix Factorization \citep{Ludewig2018}, \bpr: Bayesian Personalized Ranking \citep{Rendle2009}, \fism: Factored Item Similarity Models \citep{kabbur13fism}, \fossil: FactOrized Sequential Prediction with Item SImilarity ModeLs \citep{he16fusing}.}
Their findings showed that
\begin{enumerate*}[label=\textit{(\roman*)}]
\item these approaches either are generally not competitive in this setting or only lead to competitive results in a few specific cases and
\item that nearest neighbor recommenders outperform them in terms of prediction accuracy.
\end{enumerate*}

\paragraph{Session-based Recommendation Approaches}
While there exist some earlier works on session-based recommendation, e.g., in the context of website navigation support and e-commerce \citep{Mobasher2002,shani05mdp}, research on this topic started to considerable grow only in the mid-2010s. These developments were particularly spurred by the release of datasets in the context of machine learning competitions, e.g., at ACM RecSys 2015. At around the same time, deep learning methods were increasingly applied for recommendation problems in general. The first deep learning approach to session-based recommendation was \gru \citep{Hidasi2016GRU}, which is based on Recurrent Neural Networks. Later on, various other types of neural architectures were explored, including attention mechanisms. convolutional neural networks, graph neural networks or hybrid architectures, see \cite{wang2020survey} for an overview.

Recent work however indicates that in many cases much simpler methods can achieve similar or even higher performance levels than today's deep learning models. Most recently, Ludewig et al.~\cite{ludewiglatifiumuai2020} benchmarked several of the mentioned neural methods against conceptually simpler session-based algorithms which, for example, rely on nearest-neighbor techniques. Quite interestingly, their analyses and similar previous works \citep{Gargetal2019,Ludewig2018} not only show the strong performance of conceptually simple techniques, but also revealed that two of the earlier neural methods, \gru and \narm, often perform better than more recent complex techniques.  In the performance comparison in this present work on session-aware recommendation, we include several techniques for session-based recommendation as baselines. This allows us to assess the added value of considering long-term preference information compared to a situation where such information is not available or not leveraged.

\paragraph{Session-aware Recommendation Approaches}
The literature on session-aware recommendation is still quite sparse. An early approach is discussed in \citep{JannachLudewigLerche2017umuai}. One main goal of their work was to understand the relative importance of short-term user intents when visiting an e-commerce site compared to the long-term preference model. Their analyses, which were based on a large but private e-commerce dataset, emphasized the importance of considering the most recent observed user behavior when recommending. Furthermore, it also turned out that \emph{reminding} users of items that they have viewed before can be beneficial, both in terms of accuracy measures and business metrics.

While the work in \cite{JannachLudewigLerche2017umuai} relied on deep learning for the final predictions in one of their models, the core of the proposed technical approach was based on feature engineering and the use of side information about the items to recommend. One of the earliest ``pure'' deep learning techniques for session-aware recommendation was proposed by \citep{quadranahgru2017}. Technically, the authors based their work on \gru, and they used a second, parallel GRU layer to model information across sessions, resulting in a model called \hgru. Their analyses showed that incorporating long-term preferences can be beneficial, i.e., \hgru was outperforming an early version of \gru in their experiments.

In the same year, Ruocco et al.~\cite{Ruocco-etal:17} proposed the \iirnn model. Like \hgru, this model uses an RNN architecture and extends a session-based technique to model inter-session and intra-session information. Like in the case of \hgru, the authors investigate the value of considering long-term preference information by comparing their method to session-based techniques.
RNNs were later on also used in the \nsar model \citep{Phuong2018} to encode session patterns in combination with user embeddings to represent long-term user preferences across sessions. In their experiments, the authors not only compare their model to session-based techniques, but also to \hgru as a representative of a session-aware approach.

A number of neural architectures other than RNNs were proposed in recent years. Hu et al.~\cite{Hu-etal:18}, for example, combine the inter-session and intra-session context with a joint context encoder for item prediction in the \ncsf approach. In the \shan model \citep{Ying-etal:18}, in contrast, the authors  leverage a two-layer hierarchical attention network to model short-term and long-term user interests.
In the \swiwo \cite{Hu-etal:17} approach,  the authors were inspired by language modeling approaches like \emph{word2vec}, 
where the underlying idea 
is that items can be seen as words, hence, predicting a relevant word based on context information is equivalent to recommending a relevant item in an ongoing session.
Finally, Cai and Hu \cite{Cai2018} proposed the \samr method, which leverages a topic-based probabilistic model to define the users' listening behavior.

Unlike most previous works, which compare a newly-proposed session-aware model with previous session-based ones, our work compares session-aware methods against each other. Furthermore, we benchmark several of these recent session-aware methods with
\begin{enumerate*}[label=\textit{(\roman*)}]
\item existing session-based techniques and
\item \textcolor{orange}{ extended versions of them that also consider long-term preference information. }
\end{enumerate*}
We describe our research methodology next.

\section{Research Methodology}
\label{sec:experiment-setup}
In this section, we describe which algorithms we selected for inclusion in our comparison. Moreover, we provide details about the experimental configuration in terms of the evaluation protocol and the used datasets. As mentioned, all datasets and the code used in the experiments are shared online to ensure reproducibility.

\subsection{Compared Algorithms}
In our experiments, we compare neural session-aware algorithms with a number of baselines. Details about the algorithms are provided next.

\subsubsection{Neural Session-Aware Algorithms}
\textcolor{orange}{We identified five recent neural approaches, which we integrated into our evaluation framework using the source provided by the authors: \hgru, \iirnn, \shan, \ncsf, and \nsar.
}

\begin{itemize}
  \item \hgru: This method \citep{quadranahgru2017} is based on the \gru algorithm. To model the interactions of a user within a session, it utilizes RNNs based on a single GRU layer. By adding an extra GRU layer, it models information across user sessions. The user-level GRU initializes the hidden state of the session-level GRU and optionally propagates the user representation in the input to the session-level GRU to personalize its recommendations.
  \item \iirnn: This method \citep{Ruocco-etal:17} extends a session-based recommender built on RNN, called intra-session RNN, by using a second RNN that is called inter-session RNN. The intra-session RNN is the same as in the \gru model. The inter-session RNN learns from the user's recent sessions and feeds the information to the intra-session RNN. At the beginning of every session, the final output of the inter-session RNN initializes the hidden state of the intra-session RNN.
  \item \shan: This model \citep{Ying-etal:18} uses a two-layer hierarchical attention network to learn a hybrid representation for each user that combines the long-term and short-term preferences. It first embeds sparse user and item inputs into low-dimensional dense vectors. The long-term user representation is a weighted sum over the embeddings of items in the long-term item set. By learning the weights, the first attention layer learns user long-term preferences. The second attention layer returns the final user representation by combining the long-term user model and the embeddings of the items in the short-term item set.
  \item  \ncsf:  This session-aware neural method \citep{Hu-etal:18} has three components:
    \begin{enumerate*}[label=\textit{(\roman*)}]
    \item the historical session encoder to represent the inter-session context,
    \item the current session encoder to represent the intra-session context, and
     \item the joint context encoder to integrate the information of the intra-session context and the inter-session context for item prediction.
     \end{enumerate*}
  \item  \nsar: 
\textcolor{orange}{This method \citep{NSAR2019} utilizes
RNNs to encode session patterns (short-term user preferences) and user embeddings to represent long-term user preferences across sessions. It supports different ways of integrating user long-term preferences with the session patterns, where 
user embeddings can either be integrated with the input or the output of the session RNNs. Moreover, with the help of a gating mechanism, the contribution of each component can be fixed or adaptive.}
\end{itemize}

To avoid a bias in the algorithm selection, we applied the following procedure to identify algorithms for inclusion in our experiments. An initial set of candidate works was retrieved through a search on Google Scholar using search terms like ``session-aware recommendation'' or ``personalized session-based recommendation''. We inspected the returned results to see if the papers fulfilled our inclusion criteria. Besides being actually a work on session-aware recommendation according to the above definition, we required that the source code of the method was publicly available and could be integrated into our Python-based evaluation framework\footnote{
\textcolor{blue}{For example, the code used in \cite{Cai2018} is not publicly available, and the method in \cite{Hu-etal:17} is written in MATLAB.}}.
Moreover, we only considered papers that had undergone a peer review process, i.e., we did not include non-reviewed preprints. Finally, we included only works that did not consider side information about the items. As a result of this last constraint, we did not include works like \cite{Symeonidis-etal:20} for the domain of news recommendation.

In Table \ref{tab:used-baselines-neural-approaches}, we show to which baselines the selected neural approaches were compared in their original publications. Note that in this table only the last two rows, \hgru and \swiwo represent session-aware techniques. Our analysis furthermore shows that researchers use a variety of baselines in their experiments, which contributes to the difficulty of understanding what represents the state-of-the-art.

\begin{table}[p]
\label{tab:used-baselines-neural-approaches}
\caption{
Overview of the baseline techniques that each neural session-aware approach was originally compared to. The methods are ordered chronologically by the date of publication. The marks (\xmark) indicate which baselines were used in the comparison.
}
\footnotesize
\centering
\resizebox{0.8\textwidth}{!}{%
\begin{tabular}{llccccc}
\toprule
 & \multicolumn{1}{l}{} & \multicolumn{5}{c}{Method}         \\
 & \multicolumn{1}{l}{} & \multicolumn{1}{l}{\hgru} & \multicolumn{1}{l}{\iirnn} & \multicolumn{1}{l}{\shan} & \multicolumn{1}{l}{\ncsf} & \multicolumn{1}{l}{\nsar} \\ \toprule
\multirow{14}{*}{\begin{sideways}Baseline\end{sideways}} & \mr &           & \xmark         & & & \\
 & \pop  &  & \xmark  & \xmark & \xmark & \\
 & \ppop & \xmark  & & & & \\
  & \iknn & \xmark & \xmark & & \xmark & \\
  & \bpr    & & \xmark & \xmark & & \xmark \\
  & \fpmc & & & \xmark & \xmark & \\
  & \fossil          & & & \xmark & & \\
  & \gru & \xmark & \xmark & & \xmark & \xmark \\
  & \grutwo          & & & & \xmark & \xmark \\
  & \bprgrutwo       & & & & & \xmark \\
  & \hrm & & & \xmark & & \\
  & \caser  & & & & & \xmark \\
  & \hgru & & & & & \xmark \\
  & \swiwo           & & & & \xmark & \\ \midrule
  \multicolumn{7}{l}{\begin{tabular}[c]{@{}l@{}}\mr: most recent interacted item; \pop: most popular item in the dataset; \\ \ppop: most popular item for the user; \iknn: Item-based kNN \citep{Hidasi2016GRU}; \\ \bpr: Bayesian Personalized Ranking \citep{Rendle2009}; \fpmc: Factorized \\ Personalized Markov Chains \citep{Rendle2010}; \fossil: FactOrized Sequential \\ Prediction with Item SImilarity ModeLs \citep{he16fusing}; \gru: \citep{Hidasi2016GRU}; \\ \grutwo: the improved version of the \gru model \citep{Hidasi:2018}; \\ \hrm: Hierarchical Representation Model \citep{Want-etal:15}; \caser: Convolutional \\ Sequence Embedding Recommendation Model \citep{Tang:18}; \bprgrutwo: \\ a baseline proposed in \cite{NSAR2019} that merges the rankings returned by \\ \bpr and \grutwo following the proposed framework in \cite{Jannachetal2015} for \\ session-aware setting; \hgru: \citep{quadranahgru2017}; \swiwo: \citep{Hu-etal:17}.\end{tabular}}
  \\ \bottomrule
\end{tabular}}
\end{table}

\subsubsection{Neural and Non-Neural Session-based Baselines}
We selected the baselines according to the results of \cite{ludewiglatifiumuai2020}.
Note that while \gru and \narm are not the most recent neural methods, the analysis in \citep{ludewiglatifiumuai2020} showed that they are highly competitive among the neural models for different datasets.

\begin{itemize}
  \item \gru:  This neural model \citep{Hidasi2016GRU} employs RNNs based on Gated Recurrent Units (GRU) for the session-based recommendation task. It introduces several modifications, including a ranking loss function, to classic RNNs to adapt it for the recommendation task in the session-based setting. The authors later on improved the model with an alternative loss function and by applying further refinements, \citep{Hidasi:2018}. We include the latest version of \gru in our experiments.
  \item \narm:  This model \citep{Lietal2017} extends \gru. It utilizes a hybrid encoder with an attention mechanism to model the sequential behavior of users and capture their main intent of the current session. 
      A bi-linear matching scheme is used to compute the recommendation scores of each candidate item based on a unified session representation.
  \item \sr: The \emph{Sequential Rules} method, proposed in \citep{Ludewig2018}, extends the simple \emph{Association Rules} technique (also from \citep{Ludewig2018}) and counts pairwise item co-occurrences in the training sessions. It considers the order of the items in a session as well as the distance between them when scoring the items. 
  \item  \vsknn: This nearest-neighbor baseline for session-based recommendation was proposed in \citep{Ludewig2018}, and it is based on the \sknn method \citep{JannachLudewig2017}. It first finds past sessions that contain the same items as the current session. The recommendation list is then generated using items that occur in the most similar sessions. This method considers the order of the items while computing both session similarities and item scores. Moreover, it applies  the Inverse-Document-Frequency (IDF) weighting scheme to put more emphasis on less popular items.
  \item \stan: The \emph{Sequence and Time Aware Neighborhood} method was proposed in \citep{Gargetal2019}. It improves \sknn by considering three additional factors for making recommendations:
    \begin{enumerate*}[label=\textit{(\roman*)}]
    \item the recency of an item in the current session,
    \item the recency of a past session w.r.t.~the current session, and
    \item the distance of a recommendable item w.r.t.~a shared item in the neighboring sessions.
    \end{enumerate*}
  \item \vstan: This nearest-neighbor session-based recommendation algorithm combines all the extensions to \sknn from \stan and \vsknn in a single approach; proposed in \citep{ludewiglatifiumuai2020}.

\end{itemize}

\subsubsection{Extensions of Session-based Baselines}
\label{subsec:extensions}
We experimented with three simple ways of extending session-based algorithms in a way that they consider past preference information. 

\begin{itemize}
  \item \emph{\textsc{Extend} -- Extending the current session with recent interactions}: In case there is little information available about the ongoing session, e.g., when only a few first clicks are recorded, we extend the current session with interactions that we observed in the previous sessions of the user.
  \item \emph{\textsc{Boost} -- Emphasizing previously seen items:} In some domains, repeated interactions with already seen or consumed items are not uncommon. We apply a simple ``boosting'' approach to slightly increase the scores computed by an underlying algorithm in case an item has appeared previously in the interaction history.
  \item \emph{\textsc{Remind} -- Applying reminding techniques:} In \citep{JannachLudewigLerche2017umuai}, the authors proposed a number of ``reminder'' techniques to emphasize items the user has seen before. The general approach is to determine and score a set of candidate items that the current user has interacted with in the past. We considered \textcolor{blue}{different} strategies that were inspired by \citep{JannachLudewigLerche2017umuai} in our evaluations.
\end{itemize}

\paragraph{\textsc{Extend}} We implemented this strategy as follows. First, we choose a value for the ``desired session length'' $d$, which is a hyperparameter to be determined on the validation set. In case an ongoing session has fewer interactions than $d$, we extend the current session with previous interactions from the same user until the session length $d$ is reached or no more previous interactions exist. The extension is done by simply prepending the elements of previous interactions to the current session in the order they appeared in the log.

\paragraph{\textsc{Boost}} This simple approach can in principle be applied to any algorithm that returns scores.
Methods like \sr and \vsknn, for example, return scores based on item co-occurrences and the positions of the co-occurring items, as described in \cite{Ludewig2018}. In our experiments, we used a hyperparameter $b$ as a boost factor. Technically, we look up each item that is recommended by the underlying method
and check if it occurred in the interaction history of the current user at least once. In case the item appeared previously in the history, we increase
the original score by $b$~\%.

\paragraph{\textsc{Remind}}
\color{blue}
Different reminding strategies were proposed in \citep{JannachLudewigLerche2017umuai}. In our experiments, we tested a number of alternative ways to select and score items to consider as reminders.
In all of them, the reminder list is created by adding items of the user's last $p$ sessions, which is a hyperparameter to be determined on the validation set.
The assumption is that very old sessions at some stage might become irrelevant and should not be considered anymore.
The following reminder strategies are the best performing ones according to our experiments.

\emph{Interaction Recency}. In this strategy, we use a decay function to score reminder items, i.e., items in the interaction history of the user, based on the time of the user's last interaction with them: 
\begin{equation}
\textrm{IRecScore}(i) = T_c/(T_c - T_i)
\end{equation}

Here, $T_c$ is the timestamp of the current session, and $T_i$ is the timestamp of the last interaction of the user with the item $i$.

\emph{Session Similarity}. In this strategy, the items of the last $P$ sessions of the user are scored based on the similarity score(s) of the current session and the previous session(s) that they belong to. This strategy is based on a nearest-neighbor recommendation method to calculate the similarity scores between sessions:

\begin{equation}
\textrm{SSimScore}(i) = \sum_{p=1}^{P} Sim(S_c,S_p) . 1_p(i)
\end{equation}

Here, $S_c$ denotes the current session, and $S_p$ denotes a session in the set of the last $P$ past sessions of the user.
\emph{Sim} is the similarity function of the nearest-neighbor algorithm, and the indicator function $1_p(i)$ returns 1 if the session $p$ contains item $i$ and 0 otherwise.

In our present work, we used a \emph{hybrid} approach to combine aspects of interaction recency (IRec), session similarity (SSim), and the item's relevance score (RelScore), as determined by a recommendation algorithm, in a weighted approach. The overall ranking score for an item $i$ is therefore defined as
\begin{equation}
\textrm{OverallScore}(i) = W1 \times \textrm{RelScore}(i) + W2 \times \textrm{IRecScore}(i) + W3 \times \textrm{SSimScore}(i)
\end{equation}

Here, W1, W2, and W3 are hyperparameters to tune, and the individual scores are normalized before they are combined.
Note that for algorithms that are not based on nearest-neighbors, we do not compute a \textrm{SSimScore} value, and we thus set W3 to 0.

\color{black}

\subsection{Datasets}
We conducted our evaluations on four public datasets from three different domains: e-commerce, social networks and music.

\begin{itemize}
\item\textit{\retailrocket}: A dataset published by the e-commerce personalization company \textit{Retail Rocket}. It covers user interactions with a real-world e-commerce website over 4.5 months.

\item\textit{\xing}: A dataset published in the context of the ACM RecSys 2016 Challenge that contains interactions of job postings on a career-oriented social networking site, \textit{XING}, from about three months. It contains a fraction of \textit{XING} users and job postings, and the data is enriched with artificial users for privacy reasons. The interactions include four types of actions: \textit{click, bookmark, reply, and delete}. We filtered out all interactions of type \textit{delete} from the dataset for our evaluation, as done in \citep{quadranahgru2017}, because they are considered as negative interactions.
\item\textcolor{blue}{\textit{\cosmetics}: An e-commerce dataset containing the event history of a cosmetics shop for five months.\footnote{\url{https://www.kaggle.com/mkechinov/ecommerce-events-history-in-cosmetics-shop}}
The interactions include four types of actions: \textit{click, view, add-to-cart, remove-from-cart, purchase}. Since the methods examined in our work are not designed to consider multiple types of actions, we only used the interactions of one type (\textit{item views}) for our evaluation, as was done also in various previous works \cite{Hidasi2016GRU, Lietal2017}.} \textcolor{blue}{Furthermore, we randomly sampled 10\% of the users of this large dataset because of scalability issues of some of the neural methods.}
\item\textit{\lastfm}: A music dataset that contains the entire listening history of almost 1,000 users during five years. The dataset was retrieved from the online music service \textit{Last.fm}\footnote{\url{https://www.last.fm/}}.

\end{itemize}

For each dataset, we first partitioned the log into sessions by applying a commonly used 30-minute user inactivity threshold\footnote{The \cosmetics dataset already contains session IDs, but we noticed that there were large inactivity thresholds and some of the original sessions spanned several days.}. We kept multiple interactions with the same item in one session because repeated recommendations can help as reminders \citep{JannachLudewigLerche2017umuai,LercheJannachLudewig2016}. Many previous works on session-aware recommendation use a single training-test split of the whole dataset or a sample of it for evaluation. Evaluating on only one split of data is risky because of possible random effects.
We therefore split each dataset into five contiguous subsets by time and averaged the results across slices as done in \cite{ludewiglatifiumuai2020}. \textcolor{blue}{To have about the same number of events for each slice, we skipped the first 500 days of the \lastfm dataset.}

Following common practice in the field, we then further pre-processed each slice as follows. We first removed items with less than five interactions in the slice. Then, we removed sessions that contain only one event. For the \lastfm dataset, we also removed sessions with more than 20 events\footnote{The dataset contains a number of very long sessions with dozens of listening events, and the probability that users \emph{actively} listened to tracks for many hours seems low. Therefore, we only considered the first 20 elements, which corresponds to a listening session about 1.5 hours for the case of pop music, see also \citep{Phuong2018,Ruocco-etal:17} for similar approaches.}. Finally, we filtered out users with less than three sessions. Table \ref{tab:dataset-characteristics} shows the average characteristics of the slices for each dataset after the pre-preprocessing phase.

\begin{table}[ht]
	\centering
    \footnotesize
	\caption{Characteristics of the datasets. The values are averaged over all five slices.}
	\label{tab:dataset-characteristics}
	\resizebox{\textwidth}{!}{%
	\begin{tabular}{lrrrrrrr}
		\toprule
    	&   & Events    & Users     & Sessions      & Items     & Sessions per User   & Actions per Session \\ \midrule
    	\retailrocket	&   & 45,378    & 1,400  & 7,198    & 10,424   &  5.15   & 6.28     \\
    	\xing  &    & 333,625   & 13,533    & 59,318    & 61,006    & 4.38     & 5.62    \\
    	\cosmetics &   & 81,159   & 1,821    & 9,826   &  17,790  & 5.39    & 8.24 \\
    	  \lastfm &   & 750,276   & 658    & 94,818   & 107,134   & 144.61    & 7.92 \\
		\bottomrule
	\end{tabular}}
\color{black}
\end{table}

\subsection{Evaluation Protocol and Metrics}

\paragraph{Creation of Training, Validation, and Test Splits} Since we are given user-IDs for the sessions, we are able to apply a user-wise data splitting approach. Specifically, like in \citep{Guo2020,Phuong2018,NSAR2019,quadranahgru2017,Xiao2019}, we use the last session of each user as test data. The second-to-last session is used as validation data to tune the parameters in our experiments, see also \citep{Phuong2018,NSAR2019,Xiao2019}. The remaining sessions are considered as training data. This splitting approach allows us to assess the performance both for users with shorter and longer interaction histories.
We finally filter out the items from the validation and test sets of each of the five slices that did not appear in the training set of that slice.

\paragraph{Target Item Selection}
We apply a procedure that is commonly used also in session-based recommendation, e.g., in \citep{Hidasi2016GRU} and many other works. Specifically, we iteratively reveal each item after the other in the test session and do the evaluation after each item\footnote{
In the implementation of the \ncsf method, we found that the authors use a \emph{context window}, which includes items \emph{before and after} a given target item $t$ to make the prediction. In reality, however, we cannot know which items will appear after $t$. As we could not resolve this issue with the authors, we used an implementation that only uses items appearing \emph{before} the target item as the context.}.

We apply this approach as it reflects the most realistic user behavior in a session.  Following previous works \citep{Ludewig2018, ludewigetal2019, ludewiglatifiumuai2020}, we consider two evaluation scenarios. In one case, we only consider the immediate next item in the test session as a ground truth. In the other, more realistic case, all upcoming items in the test session are considered relevant.

\paragraph{Accuracy Metrics} We use standard classification and ranking measures to evaluate the performance of the recommendation algorithms. We measure the performance in two different ways, as done in \citep{ludewiglatifiumuai2020}, according to the target item selection approach. First, when we consider only the immediate next item as the target item, the used metrics are the Hit Rate (HR) and Mean Reciprocal Rank (MRR). Second, when all items of the current session are assumed to be relevant to the user, we consider all the remaining items of an ongoing session as the target item. In this case, the used accuracy metrics are Precision, Recall, and Mean Average Precision (MAP).

\paragraph{Coverage and Popularity Bias Metrics} It is well known that factors other than accuracy can impact the effectiveness of a recommendation algorithm in practice \citep{shani2011evaluating}.
In this work, we consider \emph{coverage} and the tendency of an algorithm to focus on popular items (\emph{popularity bias}) as relevant factors. \textit{Coverage} tells us how many items actually ever appear in the top-n lists of users. This measure, which is also known as ``aggregate diversity'' \citep{Adomavicius2011}, gives us some indication of how strongly personalized the recommendations of an algorithm are. A strong popularity bias, on the other hand, indicates that an algorithm is not focusing much on the long-tail of the item catalog, which, however, can be desirable in practice.  We calculate the \textit{popularity bias} as done in \citep{Ludewig2018}. Specifically, we average the popularity scores of all recommended items. These popularity scores are computed by counting how often each item appears in the training set. To bound their values between 0 and 1, we apply min-max normalization.

\color{blue}
\paragraph{Computational Complexity} Training deep learning models is often considered computationally demanding. Nearest-neighbor techniques, on the other hand, have no model-building phase, but the search for neighbors can, depending on the implementation, be computationally complex at run time. In our experiments, we therefore use the recency-based sampling approach proposed in \cite{JannachLudewig2017} for all session-based nearest-neighbor approaches.
To compare the computational complexity of the various methods, we measured the time that is needed by each algorithm, both for the training and the prediction phases. All these measurements were made on the same physical machine, which was exclusively used for the time measurements. The machine is equipped with a Nvidia Titan Xp GPU and an Intel i7-3820 CPU. 
The neural models used the GPU whereas the non-neural techniques only used the CPU.
\color{black}

\paragraph{Hyperparameter Optimization} To obtain reliable results, we systematically and automatically tuned the hyperparameters for each algorithm and dataset. Technically, we applied a random hyperparameter optimization procedure with 100 iterations to optimize MRR@20 as done, e.g., in \citep{Ludewig2018}\footnote{We also tried MAP@20 as the optimization target for some approaches, but this did not lead to a different ranking of the algorithms in terms of accuracy.}. For \narm, we however only ran 50 iterations as this method has a smaller set of hyperparameters. For \shan, we only ran 9 hyperparameter configurations since they cover all possible value combinations according to the original paper\footnote{We unsuccessfully contacted the authors regarding the hyperparameter spaces. 
}. For each dataset, we used the slice with the most number of events to tune hyperparameters.

\section{Results}
\label{sec:results}

Tables \ref{tab:results-knn-session-aware-retail}--\ref{tab:results-knn-session-aware-lastfm} show the results of our performance comparison of neural and non-neural methods, ordered by the values obtained for the MAP@20 metric.
Here,  we correspondingly report the values obtained by applying a cut-off threshold of 20. We performed additional experiments using alternative cut-off lengths (5 and 10). The rankings of the algorithms for those other cut-off values were generally in line with those observed at list length 20.
Non-neural methods are highlighted with a light gray background in the tables. Neural session-\emph{based} methods have a gray background. Session-\emph{aware} techniques finally have a dark gray background.

Note that we do not report all possible combinations of the proposed extensions discussed in Section \ref{subsec:extensions} for the sake of conciseness. \textcolor{blue}{We use the following naming scheme for the different algorithm variants.}
\textcolor{blue}{
\begin{itemize}
  \item \textsc{Remind}: We denote algorithm variants that were extended with the reminder technique with the postfix ``\textsc{\_r}''. Note that it is not meaningful to incorporate the reminder extensions to session-\emph{aware} methods, as these models should already be able to implicitly leverage the long-term preference information.
  \item \textsc{Extend} and \textsc{Boost}: We report the effects of these extensions for the non-neural methods\footnote{\textcolor{blue}{In principle, these mechanisms can also be applied to the neural session-based methods. Initial experiments with the \textsc{Extend} extension for \gru and \narm however did not lead to performance gains. Note also that the adaptation of existing neural session-based methods is not the main focus of our work.}}, and we denote the extended method by appending the postfix ``\textsc{\_e}'' (extend) and ``\textsc{\_b}'' (boost) to the algorithm name, e.g., \usr, \uvsknn. These extensions can also be combined with the reminder technique, e.g., \uvsknnreminder. For the \sr method, only boosting was applied because extending the session is not applicable for this algorithm, which by design only considers the last interaction in a session.
\end{itemize}
}
\color{blue}
For the sake of clarity, for each model and dataset, we report  
    \begin{enumerate*}[label=\textit{(\roman*)}]
    \item   the combination of the extensions that led to the best performance according to MAP@20 and
    \item the results of the original models without extensions.
    \end{enumerate*}
    In case the original model had better performance than the extended ones, we only report
    the results for the original model.
\color{black}

\definecolor{1}{HTML}{F5F5F5}
\definecolor{2}{HTML}{DCDCDC}
\definecolor{3}{HTML}{C0C0C0}

\begin{table}[!p]
	\caption[Results of the performance comparison]{Results of the performance comparison on the \retailrocket dataset with the focus on the comparison of \textbf{simple (non-neural)} methods and  \textbf{neural session-aware} ones. The best results for each metric are highlighted in bold font. The next best results for algorithms from the other category (either simple methods or session-aware ones) are underlined. \colorbox{1}{Non-neural} methods are highlighted in light gray, \colorbox{2}{session-based} ones in gray and \colorbox{3}{neural session-aware} ones in dark gray.}
	\label{tab:results-knn-session-aware-retail}
	\centering
	\resizebox{\textwidth}{!}{%
	\begin{tabular}{lrrr|rr|rr}
		\toprule
		Metrics   & MAP@20     & P@20   & R@20   & HR@20  & MRR@20 & COV@20 & POP@20 \\
		\midrule
		\multicolumn{8}{c}{\retailrocket}                             \\ \midrule
\cellcolor{1}\staner  &  \onesign{\best{0.0350}} &  \best{0.0759} &  0.5214 &  0.7043 &  0.4411 &  0.8374 &  0.0576 \\
\cellcolor{1}\vstanebr       &  \onesign{\best{0.0350}} &  0.0758 &  \onesign{\best{0.5215}} &  \onesign{\best{0.7062}} &  \onesign{\best{0.4432}} &  0.8525 &  0.0553 \\
\cellcolor{1}\uvsknnreminder         &  0.0343 &  0.0750 &  0.5097 &  0.6938 &  0.4155 &  \onesign{\best{0.8632}} &  0.0534 \\
\cellcolor{2}\narmreminder     &  0.0320 &  0.0691 &  0.4841 &  0.6419 &  0.3850 &  0.8008 &  0.0614 \\
\cellcolor{1}\stan              &  0.0311 &  0.0680 &  0.4837 &  0.6747 &  0.4411 &  0.7818 &  0.0582 \\
\cellcolor{1}\vstan             &  0.0310 &  0.0675 &  0.4825 &  0.6745 &  0.4397 &  0.7947 &  0.0580 \\
\cellcolor{1}\vsknn            &  0.0303 &  0.0669 &  0.4646 &  0.6476 &  0.4164 &  0.8055 &  \scnd{0.0469} \\
\cellcolor{1}\usrreminder             &  0.0301 &  0.0662 &  0.4547 &  0.6018 &  0.3666 &  0.8063 &  0.0552 \\
\cellcolor{2}\grureminder &  0.0292 &  0.0621 &  0.4565 &  0.6417 &  0.4305 &  0.9086 &  0.0405 \\
\cellcolor{2}\gru          &  0.0272 &  0.0576 &  0.4367 &  0.6172 &  0.4196 &  0.9059 &  0.0394 \\
\cellcolor{2}\narm              &  0.0252 &  0.0542 &  0.4086 &  0.5650 &  0.3566 &  0.7620 &  0.0622 \\
\cellcolor{3}\iirnn      &  \scnd{0.0239} &  \scnd{0.0524} &  \scnd{0.3775} &  0.5108 &  0.3190 &  \scnd{0.7709} &  0.0689 \\
\cellcolor{3}\hgru          &  0.0226 &  0.0485 &  0.3681 &  \scnd{0.5165} &  \scnd{0.3296} &  0.7502 &  \best{0.0425} \\
\cellcolor{3}\ncsf              &  0.0217 &  0.0468 &  0.3625 &  0.5042 &  0.3120 &  0.6871 &  0.0967 \\
\cellcolor{1}\sr                &  0.0213 &  0.0460 &  0.3477 &  0.4847 &  0.3265 &  0.7186 &  0.0528 \\
\cellcolor{3}\shan              &  0.0205 &  0.0451 &  0.3448 &  0.4498 &  0.2673 &  0.3406 &  0.1276 \\
\cellcolor{3}\nsar               &  0.0169 &  0.0370 &  0.2830 &  0.3702 &  0.2160 &  0.5813 &  0.0671 \\
                          \bottomrule
		
	\end{tabular}}
\end{table}

In our analysis, we focus on the performance comparison of non-neural methods and neural session-aware recommendation techniques. Therefore, in Table \ref{tab:results-knn-session-aware-retail} to Table \ref{tab:results-knn-session-aware-lastfm}, the highest obtained values among algorithms of these two families are printed in bold. Moreover, we underline the highest value that is obtained by the other family of algorithms.
Stars indicate significant differences (p$<$0.05) according to a Kruskal–Wallis test between all the models and a Wilcoxon signed-rank test between the best-performing techniques from either category (non-neural or neural session-aware recommendation methods).

\begin{table}[!p]
	\caption[Results of the performance comparison]{Results of the performance comparison on the \xing dataset with the focus on the comparison of \textbf{simple (non-neural)} methods and \textbf{neural session-aware} ones. The best results for each metric are highlighted in bold font. The next best results for algorithms from the other category (either simple methods or session-aware ones) are underlined. \colorbox{1}{Non-neural} methods are highlighted in light gray, \colorbox{2}{session-based} ones in gray and \colorbox{3}{neural session-aware} ones in dark gray.}
	\label{tab:results-knn-session-aware-xing}
	\centering
	\resizebox{\textwidth}{!}{%
	\begin{tabular}{lrrr|rr|rr}
		\toprule
		Metrics   & MAP@20     & P@20   & R@20   & HR@20  & MRR@20 & COV@20 & POP@20 \\
        \midrule

		\multicolumn{8}{c}{\xing}                                   \\ \midrule
\cellcolor{1}\vstanr           &  \onesign{\best{0.0194}} &  \onesign{\best{0.0497}} &  \onesign{\best{0.3031}} &  \onesign{\best{0.4445}} &  \onesign{\best{0.2837}} &  0.9581 &  0.0373 \\
\cellcolor{1}\stanr           &  \onesign{\best{0.0194}} &  0.0495 &  0.3027 &  0.4436 &  0.2828 &  0.9563 &  0.0395 \\
\cellcolor{1}\vsknnreminder    &  0.0182 &  0.0466 &  0.2880 &  0.4304 &  0.2691 &  \onesign{\best{0.9660}} &  0.0344 \\
\cellcolor{2}\narmreminder    &  0.0174 &  0.0448 &  0.2747 &  0.3961 &  0.2095 &  0.9400 &  0.0452 \\
\cellcolor{1}\usrreminder             &  0.0158 &  0.0413 &  0.2452 &  0.3463 &  0.1852 &  0.9459 &  0.0371 \\
\cellcolor{2}\grureminder &  0.0151 &  0.0387 &  0.2512 &  0.3959 &  0.2721 &  0.9363 &  0.0311 \\
\cellcolor{1}\vsknn             &  0.0139 &  0.0357 &  0.2312 &  0.3783 &  0.2605 &  0.9193 &  \scnd{0.0305} \\
\cellcolor{1}\vstan              &  0.0138 &  0.0353 &  0.2368 &  0.3890 &  0.2747 &  0.8996 &  0.0353 \\
\cellcolor{1}\stan              &  0.0137 &  0.0352 &  0.2367 &  0.3887 &  0.2734 &  0.8950 &  0.0386 \\
\cellcolor{2}\narm              &  0.0117 &  0.0304 &  0.2031 &  0.3320 &  0.2035 &  0.8252 &  0.0480 \\
\cellcolor{2}\gru           &  0.0113 &  0.0284 &  0.2007 &  0.3454 &  0.2653 &  0.9174 &  0.0270 \\
\cellcolor{3}\ncsf              &  \scnd{0.0101} &  \scnd{0.0262} &  \scnd{0.1800} &  \scnd{0.2982} &  \scnd{0.1706} &  0.7885 &  0.0683 \\
\cellcolor{1}\sr                &  0.0092 &  0.0238 &  0.1567 &  0.2532 &  0.1633 &  0.8279 &  0.0321 \\
\cellcolor{3}\nsar              &  0.0086 &  0.0229 &  0.1449 &  0.2013 &  0.0968 &  0.8268 &  0.0361 \\
\cellcolor{3}\hgru           &  0.0081 &  0.0203 &  0.1464 &  0.2524 &  0.1681 &  \scnd{0.8474} &  \best{0.0296} \\
\cellcolor{3}\iirnn            &  0.0072 &  0.0185 &  0.1274 &  0.2046 &  0.1254 &  0.8387 &  0.0484 \\
\cellcolor{3}\shan              &  0.0051 &  0.0151 &  0.0932 &  0.1231 &  0.0503 &  0.2673 &  0.1329 \\
		\bottomrule
		
	\end{tabular}}
\end{table}

\begin{table}[!p]
	\caption[Results of the performance comparison]{Results of the performance comparison on the \cosmetics dataset with the focus on the comparison of \textbf{simple (non-neural)} methods and \textbf{neural session-aware} ones. The best results for each metric are highlighted in bold font. The next best results for algorithms from the other category (either simple methods or session-aware ones) are underlined. \colorbox{1}{Non-neural} methods are highlighted in light gray, \colorbox{2}{session-based} ones in gray and \colorbox{3}{neural session-aware} ones in dark gray.}
	\label{tab:results-knn-session-aware-cosmetics}
	\centering
	\resizebox{\textwidth}{!}{%
	\begin{tabular}{lrrr|rr|rr}
		\toprule
		Metrics   & MAP@20     & P@20   & R@20   & HR@20  & MRR@20 & COV@20 & POP@20 \\
        \midrule

		\multicolumn{8}{c}{\cosmetics}                                   \\ \midrule
	\cellcolor{1}\stanebr  & \sign{\best{0.0212}} & 0.0741 & \sign{\best{0.2819}} & \sign{\best{0.4270}} & 0.1741 & 0.9585 & 0.0708 \\
	\cellcolor{1}\vstanebr & 0.0210 & \sign{\best{0.0751}} & 0.2766 & 0.3959 & 0.1682 & 0.9512 & 0.0606 \\
	\cellcolor{1}\uvsknnreminder & 0.0209 & 0.0749 & 0.2744 & 0.4128 & 0.1683 & \sign{\best{0.9714}} & 0.0601 \\
	\cellcolor{2}\grureminder & 0.0176 & 0.0617 & 0.2410 & 0.3808 & 0.1575 & 0.9114 & 0.0473 \\
	\cellcolor{2}\narmreminder    & 0.0175 & 0.0644 & 0.2337 & 0.3273 & 0.1223 & 0.9130 & 0.0674 \\
	\cellcolor{1}\vsknn       & 0.0175 & 0.0628 & 0.2405 & 0.3870 & 0.1665 & 0.9626 & \scnd{0.0562} \\
	\cellcolor{1}\stan        & 0.0173 & 0.0619 & 0.2494 & 0.4035 & 0.1760 & 0.9419 & 0.0655 \\
	\cellcolor{1}\vstan      & 0.0172 & 0.0615 & 0.2490 & 0.4041 & \sign{\best{0.1765}} & 0.9449 & 0.0648 \\
	\cellcolor{1}\usrreminder     & 0.0170 & 0.0623 & 0.2286 & 0.3386 & 0.1354 & 0.9528 & 0.0613 \\
	\cellcolor{2}\gru     & 0.0143 & 0.0504 & 0.2083 & 0.3383 & 0.1417 & 0.8811 & 0.0449 \\
	\cellcolor{3}\ncsf       & \scnd{0.0133} & \scnd{0.0489} & \scnd{0.1903} & \scnd{0.2969} & 0.1099 & 0.6973 & 0.1043 \\
	\cellcolor{2}\narm        & 0.0129 & 0.0473 & 0.1891 & 0.2970 & 0.1175 & 0.8566 & 0.0694 \\
	\cellcolor{3}\hgru   & 0.0123 & 0.0442 & 0.1797 & \scnd{0.2969} & \scnd{0.1198} & \scnd{0.9332} & \best{0.0468} \\
	\cellcolor{3}\iirnn    &  0.0123 &  0.0458 &  0.1761 &  0.2825 &  0.1119 &  0.8909 &  0.0778 \\
	\cellcolor{1}\sr         & 0.0111 & 0.0411 & 0.1654 & 0.2755 & 0.1161 & 0.9179 & 0.0574 \\
	\cellcolor{3}\nsar       & 0.0081 & 0.0324 & 0.1217 & 0.1845 & 0.0617 & 0.7748 & 0.0680 \\
	\cellcolor{3}\shan        & 0.0054 & 0.0226 & 0.0898 & 0.1225 & 0.0434 & 0.2166 & 0.1985 \\
	\bottomrule
		
	\end{tabular}}
\end{table}

\begin{table}[!p]
	\caption[Results of the performance comparison]{Results of the performance comparison on the \lastfm dataset with the focus on the comparison of \textbf{simple (non-neural)} methods and \textbf{neural session-aware} ones. The best results for each metric are highlighted in bold font. The next best results for algorithms from the other category (either simple methods or session-aware ones) are underlined. \colorbox{1}{Non-neural} methods are highlighted in light gray, \colorbox{2}{session-based} ones in gray and \colorbox{3}{neural session-aware} ones in dark gray.}
	\label{tab:results-knn-session-aware-lastfm}
	\centering
	\resizebox{\textwidth}{!}{%
	\begin{tabular}{lrrr|rr|rr}
		\toprule
		Metrics   & MAP@20     & P@20   & R@20   & HR@20  & MRR@20 & COV@20 & POP@20 \\
		\midrule
		
		\multicolumn{8}{c}{\lastfm}                                \\ \midrule
\cellcolor{1}\uvsknn       &  \onesign{\best{0.0493}} &  \onesign{\best{0.1003}} &  \onesign{\best{0.4248}} &  \onesign{\best{0.5383}} &  0.1771 &  0.5053 &  0.0499 \\
\cellcolor{1}\vsknn          &  0.0474 &  0.0964 &  0.4104 &  0.5227 &  0.1725 &  0.5091 &  \scnd{0.0479} \\
\cellcolor{1}\vstaneb      &  0.0449 &  0.0943 &  0.4088 &  0.5291 &  0.1938 &  0.5080 &  0.0528 \\
\cellcolor{1}\stanebr       &  0.0442 &  0.0924 &  0.3972 &  0.5159 &  0.1902 &  0.4965 &  0.0564 \\
\cellcolor{1}\stan           &  0.0400 &  0.0856 &  0.3760 &  0.5012 &  0.1886 &  0.5067 &  0.0557 \\
\cellcolor{1}\vstan          &  0.0394 &  0.0852 &  0.3791 &  0.5100 &  0.1935 &  \scnd{0.5115} &  0.0555 \\
\cellcolor{1}\usr           &  0.0359 &  0.0778 &  0.3408 &  0.4622 &  \scnd{0.3487} &  0.5010 &  0.0549 \\
\cellcolor{1}\sr             &  0.0348 &  0.0765 &  0.3383 &  0.4615 &  0.3350 &  0.4982 &  0.0544 \\
\cellcolor{3}\iirnn     &  \scnd{0.0311} &  \scnd{0.0724} &  \scnd{0.3270} &  \scnd{0.4729} &  \best{0.3491} &  0.4443 &  0.0732 \\
\cellcolor{2}\gru        &  0.0307 &  0.0701 &  0.3175 &  0.4681 &  0.3342 &  0.5124 &  0.0468 \\
\cellcolor{3}\nsar           &  0.0280 &  0.0675 &  0.3044 &  0.4350 &  0.2906 &  0.4937 &  0.0483 \\
\cellcolor{2}\narm           &  0.0272 &  0.0658 &  0.3002 &  0.4510 &  0.3192 &  0.4748 &  0.0633 \\
\cellcolor{3}\ncsf           &  0.0219 &  0.0556 &  0.2639 &  0.4393 &  0.2434 &  0.4875 &  0.0712 \\
\cellcolor{3}\hgru       &  0.0208 &  0.0517 &  0.2464 &  0.4206 &  0.3167 &  \best{0.5647} &  \best{0.0404} \\
\cellcolor{3}\shan           &  0.0072 &  0.0223 &  0.0920 &  0.1149 &  0.0345 &  0.0824 &  0.1640 \\

		\bottomrule
		
	\end{tabular}}
\end{table}

\subsection{Accuracy Results}

We can summarize the accuracy results for the individual datasets as follows.

\paragraph{\retailrocket} Quite surprisingly, simple nearest-neighbor recommenders win on all the accuracy measures on this dataset, with \textcolor{blue}{\staner and \vstanebr} being the best-performing methods.
Moreover, the heuristic extensions of the session-based algorithms further improve their accuracy performance in 
\textcolor{blue}{all but one cases (MRR for \uvsknnreminder)}.

Probably even more surprising is that we find the session-\emph{aware} methods at the very end of the performance ranking. This means that they are actually outperformed by methods which do not consider any long-term preference information at all.
\textcolor{blue}{Specifically, all neural and non-neural session-based methods, except the basic \sr method, outperform all session-\emph{aware} methods for all accuracy metrics.}

The best results in the class of  neural session-\emph{aware} recommendation algorithms are achieved by \iirnn and \hgru. These two methods are, however, the earliest proposed session-aware recommendation methods in this comparison. Differently from the original paper, \hgru is not able to outperform the last version of \gru on this dataset.\footnote{Using alternative loss functions for \hgru might help improving its performance. Such modifications of the original algorithms are however not in the scope of our work.}

\paragraph{\xing} Similar patterns are also found for the \xing dataset, where
\begin{enumerate*}[label=\textit{(\roman*)}]
\item the neighborhood-based methods led to the best results,
\item the extensions resulted in performance improvements in all cases,
\item and session-\emph{aware} techniques were outperformed by all
 other methods, except the \sr method.
\end{enumerate*}
Among the session-\emph{aware} recommendation methods, this time \ncsf achieves the best accuracy results for all metrics, even though it was not originally evaluated on this dataset.

\textcolor{blue}{Note that the reminders worked generally very well on this dataset.}
Reminders for example help to improve the performance of the neural session-based techniques (\gru, \narm) to an extent that they sometimes outperform the basic nearest-neighbor techniques. However, when the extensions are also considered for the neighborhood-based techniques, their accuracy results are again much higher than those of the neural techniques.

\paragraph{\cosmetics} \textcolor{blue}{We observe similar results also on the \cosmetics dataset. Neighborhood-based methods lead to the best results for all accuracy metrics, and the extensions resulted in performance improvements in almost all cases.
Moreover, all methods except \sr and \narm outperform the session-\emph{aware} techniques.
Looking at the session-\emph{aware} methods, we notice that \ncsf has the best performance in all the accuracy metrics, except for the MRR, and \hgru achieves the best results for the Hit Rate (along with \ncsf) and the MRR.}

\paragraph{\lastfm} \textcolor{blue}{The picture for this dataset is slightly different from the others in certain respects. First, while nearest-neighbor approaches again lead to the best results for Precision, Recall, MAP, and HR, these methods are outperformed by all neural session-based and session-aware methods except \shan on the MRR.
However, none of them, except \iirnn, outperforms the simple \sr method and its extended version.
The best results are obtained by \iirnn, which is however one of the earliest session-aware methods. Note, however, that the results by \iirnn are only minimally better than \usr.
Nonetheless, further investigations are needed to understand the reasons why some methods perform very well on the MRR on this particular dataset.}

\textcolor{blue}{A second difference to the other datasets is that while applying the simple extensions \textsc{Extend} and \textsc{Boost} again proves to be beneficial, the reminder extension does not improve the performance of the original methods in some cases for this dataset.\footnote{In some cases, the optimal weights for the reminders were 0 after tuning the hyperparameters.} Therefore, we only report the results of the original neural session-based methods (\gru and \narm).}

\paragraph{Summary and Additional Observations} Table \ref{tab:results-knn-neural} summarizes the findings presented in Table \ref{tab:results-knn-session-aware-retail}--\ref{tab:results-knn-session-aware-lastfm}
by reporting the best performing approaches on the individual datasets. In Table \ref{tab:results-knn-neural}, we can see that \vsknn, \stan and \vstan with their extensions largely dominate the field across the datasets and measures, and we recommend that these methods are considered as additional baselines in future performance comparisons.
\textcolor{blue}{Only in one case a method from another category (i.e., IIRNN, an early proposed neural session-aware method) appeared among the best performing approaches.}

\textcolor{blue}{
Looking at the best models among the neural approaches, including both session-based and session-aware, in Table \ref{tab:results-neural-best}, we notice that \narmreminder and
\grureminder outperform the session-aware models on the \retailrocket, \xing, and \cosmetics datasets.
On the \lastfm dataset, where the reminder extension did not lead to any performance improvements for the neural session-based models, \iirnn is the best one.}
As a side observation, we see that the ranking of the neural algorithms is often not correlated with the publication year of the methods, i.e., newer methods are not consistently better than older ones.

\begin{table}[ht]
	\caption[Best performing approaches]{
	Best performing approaches for each dataset and each accuracy metric.
	Algorithms that were significantly better than the best performing algorithms from the other category (\textcolor{orange}{either \textbf{non-neural} or \textbf{neural session-aware}}) are marked with *.
	}
	\label{tab:results-knn-neural}
	\centering
	\scriptsize
	\begin{tabular}{lllll}
		\toprule
		Datasets & \retailrocket                    & \xing                    & \cosmetics       & \lastfm                     \\
		\midrule
		MAP@20   & \onesign{\staner/\vstanebr}  & \onesign{\vstanr/\stanr} & \onesign{\stanebr}   & \onesign{\uvsknn} \\
		P@20     & \staner                      & \onesign{\vstanr}        & \onesign{\vstanebr}  & \onesign{\uvsknn}              \\
		R@20     & \onesign{\vstanebr}          & \onesign{\vstanr}        & \onesign{\stanebr}   & \onesign{\uvsknn}             \\
		\midrule
		HR@20    & \onesign{\vstanebr}          & \onesign{\vstanr}        & \onesign{\stanebr}   & \onesign{\uvsknn}     \\
		MRR@20   & \onesign{\vstanebr}          & \onesign{\vstanr}        & \onesign{\vstan}     & \iirnn                \\
		\bottomrule
	\end{tabular}
\end{table}

\begin{table}[ht]
	\caption[Best performing approaches neural methods]{
	\textcolor{blue}{Best performing approaches for each dataset and each accuracy metric for neural methods (session-based and session-aware).}
	}
	\label{tab:results-neural-best}
	\centering
	\scriptsize
	\begin{tabular}{lllll}
		\toprule
		Datasets & \retailrocket & \xing         & \cosmetics  & \lastfm   \\
		\midrule
		MAP@20   & \narmreminder  & \narmreminder & \grureminder  & \iirnn \\
		P@20     & \narmreminder  & \narmreminder & \narmreminder  & \iirnn\\
		R@20     & \narmreminder  & \narmreminder & \grureminder   & \iirnn\\
		\midrule
		HR@20    & \narmreminder  & \narmreminder & \grureminder   & \iirnn\\
		MRR@20   & \grureminder   & \grureminder  & \grureminder   & \iirnn\\
		\bottomrule
	\end{tabular}
\end{table}

\subsection{Coverage and Popularity}

Tables \ref{tab:results-knn-session-aware-retail}--\ref{tab:results-knn-session-aware-lastfm} also report the values of \textit{coverage} and \textit{popularity bias} of the algorithms. We can make the following observations.

\paragraph{Coverage} \shan consistently has the lowest \textit{coverage} value across all datasets. In other words, it has the highest tendency to recommend the same set of items to different users. This sets the algorithm apart from all other techniques. The coverage values of most other techniques are often not too far apart, and the difference between nearest-neighbor algorithms and neural algorithms is often small.
\textcolor{blue}{However, all nearest-neighbors methods (both original and extended versions) outperform all session-\emph{aware} models for all datasets except for the \lastfm dataset, where \hgru achieves the best result.} No other consistent pattern can be found here.
What can be observed is that the achieved level of coverage and the ranking of the algorithms in this respect seems to depend on the datasets.

\paragraph{Popularity Bias} \gru and \hgru consistently have the lowest tendency to recommend popular items.
\textcolor{blue}{ In contrast, \shan exhibits the highest \textit{popularity bias} with a considerable difference. Moreover, \iirnn and \ncsf achieve higher values than all other methods across all datasets.}
The neighborhood-based approaches are often in the middle. They are therefore not generally focusing more on popular items than neural approaches. Finally, we observe that using the proposed session-aware extensions in most cases leads to a higher \textit{popularity bias}.

\begin{table}[t]
\centering
	\caption[Results of the performance comparison]{\textcolor{blue}{Running times
	on the \lastfm dataset.}}
\label{tab:times_lastfm}
\resizebox{0.6\textwidth}{!}{%
\begin{tabular}{lrr}
\toprule
      & Training Time (s) & Prediction Time (ms) \\ \midrule
\cellcolor{1}\textcolor{brown}{\sr}      & \textcolor{brown}{4.05}              & \textcolor{brown}{25.88}             \\
\cellcolor{1}\textcolor{brown}{\vsknn}   & \textcolor{brown}{1.23}              & \textcolor{brown}{169.29}            \\
\cellcolor{1}\textcolor{brown}{\stan}    & \textcolor{brown}{1.02}              & \textcolor{brown}{251.66}            \\
\cellcolor{1}\textcolor{brown}{\vstan}   & \textcolor{brown}{1.17}              & \textcolor{brown}{405.32}            \\
\cellcolor{1}\usr     & 12.58             & 37.12             \\
\cellcolor{1}\uvsknn  & 2.08              & 188.65            \\
\cellcolor{1}\stanebr & 2.81              & 466.23            \\
\cellcolor{1}\vstaneb & 1.97              & 627.66            \\
\cellcolor{2}\gru     & 101.20            & 30.18             \\
\cellcolor{2}\narm    & 3726.80           & 8.92              \\
\cellcolor{3}\hgru    & 218.71            & 27.18             \\
\cellcolor{3}\iirnn   & 7702.04           & 245.06            \\
\cellcolor{3}\ncsf    & 532.62            & 30.75             \\
\cellcolor{3}\nsar    & 1621.05           & 15.79             \\
\cellcolor{3}\shan    & 35298.70          & 30.98             \\    \bottomrule
\end{tabular}}
\end{table}

\begin{table}[t]
	\centering
	\caption[Results of the performance comparison]{\textcolor{blue}{Running times
	on the \xing dataset.}}
	\label{tab:times_xing}
	\resizebox{0.6\textwidth}{!}{%
		\begin{tabular}{lrr}
			\toprule
			& Training Time (s) & Prediction Time (ms) \\ \midrule
			\cellcolor{1}\textcolor{brown}{\sr}            & \textcolor{brown}{2.66}              & \textcolor{brown}{6.69}              \\
			\cellcolor{1}\textcolor{brown}{\vsknn}         & \textcolor{brown}{0.58}              & \textcolor{brown}{10.88}             \\
			\cellcolor{1}\textcolor{brown}{\stan}          & \textcolor{brown}{0.54}              & \textcolor{brown}{9.37}              \\
			\cellcolor{1}\textcolor{brown}{\vstan}         & \textcolor{brown}{0.58}              & \textcolor{brown}{9.85}              \\
			\cellcolor{1}\usrreminder   & 7.23              & 22.76             \\
			\cellcolor{1}\vsknnreminder & 1.09              & 25.45             \\
			\cellcolor{1}\stanr         & 1.00              & 44.26             \\
			\cellcolor{1}\vstanr        & 1.97              & 44.83             \\
			\cellcolor{2}\textcolor{brown}{\gru}           & \textcolor{brown}{53.89}             & \textcolor{brown}{10.54}             \\
			\cellcolor{2}\textcolor{brown}{\narm}          & \textcolor{brown}{1769.09}           & \textcolor{brown}{10.12}             \\
			\cellcolor{2}\grureminder   & 47.64             & 23.26             \\
			\cellcolor{2}\narmreminder  & 1762.86           & 20.37             \\
			\cellcolor{3}\hgru          & 60.54             & 10.27             \\
			\cellcolor{3}\iirnn         & 7229.03           & 250.39            \\
			\cellcolor{3}\ncsf          & 192.25            & 35.60             \\
			\cellcolor{3}\nsar          & 12562.06          & 34.81             \\
			\cellcolor{3}\shan          & 13393.82          & 35.50             \\ \bottomrule
	\end{tabular}}
\end{table}

\color{blue}
\subsection{Scalability}
For the sake of brevity, we only report the results for two datasets here and provide the results for the other datasets in the appendix.
\textcolor{blue}{We made all the measurements on the validation slice (i.e., the largest slice in terms of the number of events) for each dataset.
Table \ref{tab:times_lastfm} and Table \ref{tab:times_xing} show the results for the \lastfm and \xing, respectively.}
We selected these two datasets for the discussion because they are larger than the other two datasets and differ in some key characteristics. The \lastfm dataset for example contains the largest number of events, but only a smaller number of users. The \xing dataset, on the other hand, contains events from a larger number of users. \textcolor{brown}{The algorithms in these tables are grouped by the type of approach (non-neural, session-based neural, session-aware neural). 
We report running times of the session-based algorithms (neural and non-neural) both for the best-performing extension and the corresponding method.}

The overall results are similar across all datasets. In terms of the \emph{training} times, the non-neural models are consistently the fastest because ``training'' for such models mainly involves the initialization of data structures (nearest-neighbor techniques) or the collection of simple count statistics (\sr). Interestingly, strong differences can be observed between the neural session-\emph{based} methods \gru and \narm, with \gru being orders of magnitude faster than \narm in the training phase. The performance of the session-\emph{aware} methods, finally, exhibit a large spread. \hgru, which is based on \gru, is the fastest method here.
\textcolor{blue}{However, some session-aware methods 
can take very long\footnote{\textcolor{blue}{Note that for the \iirnn method on the \xing dataset we had to set max\_epoch=20 (instead of 100) because of its very high computational complexity on this dataset.}}, in particular the \shan method.}

Looking at the time that is needed to generate one recommendation list (\emph{prediction time}, in milliseconds), we observe that most neural methods are among the fastest techniques \textcolor{brown}{on the \lastfm dataset}, with prediction times being in the range of a few dozen milliseconds.
\textcolor{blue}{However, the running time for the neural \iirnn method, \textcolor{brown}{which is the best performing neural method on this dataset\footnote{\textcolor{brown}{Remember, however, that all nearest-neighbors methods---both the basic and extended versions---outperformed the \iirnn method on all accuracy metrics, except MRR.}}, is} much higher than for the other neural methods. For the \xing dataset, it even is the slowest among all compared methods.}
\textcolor{blue}{The running times for the simple \sr method are in the range of the neural methods.}
For the nearest-neighbor techniques, the prediction times depend on the dataset characteristics. For the \xing dataset, for example, the prediction times are in the range of the neural methods. On the other hand, for the \lastfm dataset, where,
\textcolor{blue}{on average, there is a large number of sessions}
in the history of the users, the time needed for creating one recommendation list can go up to a few hundred milliseconds.

\textcolor{brown}{Generally, the use of the proposed extensions (\textsc{Boost}, \textsc{Extend}, \textsc{Remind}) leads to increased computation times for training and predicting. Note, however, that the more efficient basic versions of the nearest-neighbor techniques already outperform all neural \textit{session-aware} methods in all cases, except one. 
}

\textcolor{brown}{Overall, as mentioned above, we observe quite a spread regarding the running times among the neural methods. A detailed theoretical analysis of the computational complexity of the underlying network architectures and individual architecture elements is however beyond the scope of our present work, which focuses on the empirical evaluation of various algorithms in terms of their prediction accuracy.}

\color{black}

\color{blue}
\section{Implications, Limitations, and Future Work}
\label{sec:implications}

\subsection{Implications and Guidelines}
Different factors may have contributed to the surprising observations made in this paper. First, we can assume that session-\emph{based} algorithms, which are often used as baselines to benchmark session-\emph{aware} ones, have improved over time. This might for example be the case for the \hgru method, which most probably used an earlier version of \gru both as a session-based baseline and as a building block for the newly proposed method.

Another potential reason may however also lie in methodological issues and problematic research practices that were observed previously not only for more traditional top-n recommendation tasks, but also for other areas of applied machine learning such as information retrieval or time-series forecasting \cite{Armstrong:2009:IDA:1645953.1646031,ferraridacrema2020tois,Makridakis2018,Yang:2019:CEH:3331184.3331340}.
According to several of these works, one main problem seems to be that researchers often invest substantial efforts to refine and fine-tune their newly proposed methods, but do not invest similar efforts to optimize the baselines to which they compare their new methods. This phenomenon can be seen as a form of a \emph{confirmation bias}, where researchers mainly seek for evidence supporting their theories and claims, but do not appropriately consider other evidences or indications. In that context, also reproducibility issues can play a role; see also the discussion of reproducibility problems in AI and empirical computer science in general \cite{Gundersen_2020,cockburn2020threats}. While researchers in recent years more frequently publicly share the code of their newly proposed models, they only very rarely share the code of the baselines or the code that was used to fine-tune all algorithms in the experimental evaluations.

Another related problem can also lie in the \emph{choice} of the baselines. In particular in recent years we observe that researchers only consider very recent baselines in their evaluations or limit themselves to neural methods \cite{ferraridacrema2020tois}. This leads to the effect that long-known or more simple methods are overlooked and not considered as competitive baselines. This, in turn, can lead to a \emph{cascade} effect, where only the most recent models are considered to represent the state-of-the-art, even though they are probably not better than what existed earlier.

A number of measures can be taken to avoid that we only observe illusions of progress in this area in the future. In terms of \emph{reproducibility}, Gundersen et al. \cite{Gundersen_Gil_Aha_2018} recently proposed a number of guidelines for reproducible research in AI in general, which in principle also apply for research in recommender systems research. Transferred to our specific problem setting, it is important that scholars make it as easy as possible for others to replicate their research. This in particular includes the publication of a number of artifacts such as
\begin{enumerate*}[label=\textit{(\roman*)}]
\item the code of the proposed model \emph{and} the baselines,
\item the used datasets, before and after any preprocessing, and the code used for pre-processing,
\item the code for running the experiments, including the code for tuning the hyper-paremeters, \item documentation and appropriate installation instructions.
\end{enumerate*}

Furthermore, in order to ensure progress, Ferrari Dacrema et al.~\cite{ferraridacrema2020tois} suggest a number of guidelines for top-n recommendation settings that are also relevant for session-aware recommendation problems. Specifically, researchers are encouraged to
\begin{enumerate*}[label=\textit{(\roman*)}]
\item include algorithms from different \emph{families} in the experimental evaluations, i.e., not only consider neural techniques,
\item optimize all baselines models in a systematic and automated way, and to
\item carefully select and document the hyperparameter ranges and the optimization strategy.
\end{enumerate*}

\subsection{Research Limitations and Future Work}

In our work so far, we performed experiments for a variety of domains---including e-commerce, music, or job recommendation \textcolor{blue}{(social media)}---using publicly available and commonly used datasets. Nonetheless, further experiments with additional datasets are needed and are part of our future work. Such experiments will help us ensure that our findings generalize to other domains and that they are not tied to specific dataset characteristics. Our experiments so far however indicate that the ranking of the \emph{family} of algorithms is quite consistent across experiments, with today's neural approaches to session-aware recommendation not being among the top-performing techniques, and non-neural techniques working well across datasets.

Nonetheless, an interesting area for future work lies in an improved understanding in which ways dataset characteristics impact the performance and ranking of different algorithms, as was done for more traditional recommendation scenarios in \cite{Adomavicious2012Impact}. Moreover, it would be interesting to analyze how sensitive the different algorithms are with respect to slight changes, including both  the characteristics of the input data and hyperparameters. In practical environments, in which datasets continuously evolve due to new items and users, algorithms that are more robust with respect to these aspects, might be preferable.
\textcolor{brown}{Finally, regarding running times---in particular for datasets where there is a larger set of past sessions per user---further performance enhancements of the nearest-neighbor methods might be achieved through additional engineering efforts.} 

Regarding the set of session-aware algorithms that were benchmarked in our work, we expect that a constant stream of new proposals will be published in the future. As such, our experimental evaluation can only represent a snapshot of the state-of-the-art in the area. Our current snapshot is furthermore limited to works where the source code was publicly available, as was done in \cite{ferraridacrema2020tois}. An additional constraint that we applied when selecting baselines was that the code had to be written in Python. In our literature search, we only identified two related works that were not written in Python, \cite{Cai2018} and \cite{Hu-etal:17}. The method proposed in \cite{Cai2018} was written in Java, but no source code was provided. The work presented in \cite{Hu-etal:17} was written in MATLAB but is only marginally relevant for our work, as it
\begin{enumerate*}[label=\textit{(\roman*)}]
\item focuses on diversity aspects and
\item provides a performance comparison only with session-based or sequential approaches, whereas our work focuses on session-aware techniques.
\end{enumerate*}

Finally, we observed that the size of the datasets used in the experiments, both in the original papers and in our work, are small compared to the amounts of data that are available in real-world deployments. The current limitations of the investigated deep learning models might therefore be a result of these dataset limitations, see also \cite{JannachdeSouzaetal2020}. With more data and in particular with data spanning longer periods of time, neural techniques might be favorable and better suited to combine long-term preference signals with short-term user intents than today's methods.

\color{black}

\section{Summary}
Our in-depth empirical investigation of five recent neural approaches to session-aware recommendation has revealed that these methods, contrary to the claims in the respective papers, are not effective at leveraging long-term preference information for improved recommendations. According to our experiments, these methods are almost consistently outperformed by methods that only consider the very last interactions of a user. 
Furthermore, our analyses showed that non-neural methods based on nearest neighbors can lead to better performance results than ones based on deep learning, as was also previously observed for session-based recommendation in \citep{ludewiglatifiumuai2020}.

\color{blue}
We see the reasons for these unexpected phenomena in methodological issues that are not limited to session-aware recommendation scenarios, in particular in the choice of the baselines in experimental evaluations or the lack of proper tuning of the baselines.
\color{black}
On a more positive note, our findings suggest that there are many opportunities for the development of better neural and non-neural methods for session-aware recommendation problems. We in particular believe that it is promising to look at repeating patterns, seasonal effects, or trends in the data. Moreover, the incorporation of side information (e.g., category information about items) as well as contextual information should help to further improve the prediction performance of new algorithms.

\section*{Acknowledgements}
We are grateful to Zhongli Filippo Hu for his contributions to the integration of the algorithms. We thank Malte Ludewig for his help and support during this research.

\bibliographystyle{plain}

\bibliography{refs}

\appendix

\section{Hyperparameter ranges and optimal values}

\begin{table}[ht]
\centering
\caption{Hyperparameter space for simple methods.}
\label{tab:space_simple}
\Huge
\def\arraystretch{0.85}
\resizebox{\textwidth}{!}{%
\Huge
\begin{tabular}{lllll}
\toprule
Algorithm/Extension                              & Hyperparameter             & Type             & Range                                    & Steps \\ \midrule
\multirow{3}{*}{\sr}    & \multirow{2}{*}{steps}     & Integer          & 2 - 15                                   & 14    \\
                                       &                            & Integer          & 20, 25, 30                                  &      \\
                                       & weighting                  & Categorical      & linear, div, quadratic, log              &       \\ \midrule
\multirow{5}{*}{\vsknn} & k                          & Integer          & 50, 100, 500, 1000, 1500                 &       \\
                                       & sample\_size               & Integer          & 500, 1000, 2500, 5000, 10000             &       \\
                                       & weighting                  & Categorical      & same, div, linear, quadratic, log        &       \\
                                       & weighting\_score           & Categorical      & same, div, linear, quadratic, log        &       \\
                                       & idf\_weighting             & Boolean/Integer & False, 1, 2, 5, 10                       &       \\ \midrule
\multirow{5}{*}{\stan}  & k                          & Integer          & 100, 200, 500, 1000, 1500, 2000          &       \\
                                       & sample\_size               & Integer          & 1000, 2500, 5000, 10000                  &       \\
                                       & lambda\_spw                & Real             & 0.00001, 0.4525, 0.905, 1.81, 3.62, 7.24 &       \\
                                       & lambda\_snh                & Real             & 2.5, 5, 10, 20, 40, 80, 100              &       \\
                                       & lambda\_inh                & Real             & 0.00001, 0.4525, 0.905, 1.81, 3.62, 7.24 &       \\ \midrule
\multirow{8}{*}{\vstan} & k                          & Integer          & 100, 200, 500, 1000, 1500, 2000          &       \\
                                       & sample\_size               & Integer          & 1000, 2500, 5000, 10000                  &       \\
                                       & similarity                 & Categorical      & cosine, vec                              &       \\
                                       & lambda\_spw                & Real             & 0.00001, 0.4525, 0.905, 1.81, 3.62, 7.24 &       \\
                                       & lambda\_snh                & Real             & 2.5, 5, 10, 20, 40, 80, 100              &       \\
                                       & lambda\_inh                & Real             & 0.00001, 0.4525, 0.905, 1.81, 3.62, 7.24 &       \\
                                       & lambda\_ipw                & Real             & 0.00001, 0.4525, 0.905, 1.81, 3.62, 7.24 &       \\
                                       & lambda\_idf                & Boolean/Integer & False, 1, 2, 5, 10                       &       \\ \midrule
BOOST                                  & boost\_own\_sessions       & Real             & 0.1-3.9                                  & 20    \\
(For \sr, \vsknn, \stan and \vstan)        &                            &                  &                                          &       \\
Tune together with other         &                            &                  &                                          &       \\
hyperparameters of the algorithm       &                            &                  &                                          &       \\ \hline
EXTEND                                 & extend\_session\_length    & Integer          & 1 - 25                                   & 25    \\
(For \vsknn, \stan and \vstan)            &                            &                  &                                          &       \\
Tune together with other               &                            &                  &                                          &       \\
hyperparameters of the algorithm       &                            &                  &                                          &       \\ \hline
                                 & \reminders & Boolean          & True                                   &     \\
REMIND                                 & \remindstrategy & Categorical          & hybrid                                   &     \\
(For \vsknn, \stan and \vstan)                                 & \remindsess & Integer          & 1 - 10                                   & 10    \\
Tune with the optimal set of            & \wbase      & Integer          & 1 - 10                                   & 10    \\
 hyperparameters of the algorithm              & \wirec      & Integer          & 0 - 9                                    & 10    \\
       & \wssim      & Integer          & 0 - 9                                    & 10    \\ \hline
REMIND                                 & \reminders & Boolean          & True                                   &     \\
(For \sr)                                 & \remindstrategy & Categorical          & hybrid                                   &     \\
Tune with the optimal set of                                  & \remindsess & Integer          & 1 - 10                                   & 10    \\
hyperparameters of the algorithm                              & \wbase      & Integer          & 1 - 10                                   & 10    \\
               & \wirec      & Integer          & 0 - 9                                    & 10    \\

               \bottomrule
\multicolumn{5}{l}{\begin{tabular}[c]{@{}l@{}}
For \stan and \vstan, we used the same hyperparameters space for lambda\_spw, lambda\_inh, and lambda\_ipw on all datasets.\\
However, ranges for these hyperparameters can be set for each dataset separately based on its mean average length of the sessions. \\
Ranges can be set as \{0.785, 1.57, 3.14, 6.28, 12.56\} for the \retailrocket dataset, \{0.7025, 1.405, 2.81, 5.62, 11.24\} for the \xing dataset, \\ \{1.03, 2.06, 4.12, 8.24, 16.48\} for the \cosmetics dataset, and \{0.99, 1.98, 3.96, 7.92, 15.84\} for the \lastfm dataset. \\ This will lead to slightly different results.
\end{tabular}}
  \\
\end{tabular}}
\end{table}

\begin{table}[ht]
\centering
\caption{Hyperparameter search space for neural methods.}
\label{tab:space_neural}
\Huge
\def\arraystretch{0.80}
\resizebox{\textwidth}{!}{%
\begin{tabular}{llllll}
\toprule
Algorithm                                     & Fixed Hyperparameter Values                                                                                                                                                                                                                          & Hyperparameter                  & Type        & Range               & Steps \\ \midrule
\multirow{7}{*}{\gru}          & \multirow{7}{*}{layer\_size = 100}                                                                                                                                                                                                                   & \multirow{2}{*}{learning\_rate} & Real        & 0.01 - 0.1          & 10    \\
                                              &                                                                                                                                                                                                                                                      &                                 & Real        & 0.1 - 0.5           & 5     \\
                                              &                                                                                                                                                                                                                                                      & momentum                        & Real        & 0 - 0.9             & 10    \\
                                              &                                                                                                                                                                                                                                                      & loss                            & Categorical & bpr-max, top1-max   &       \\
                                              &                                                                                                                                                                                                                                                      & final\_act                      & Categorical & elu-0.5, linear     &       \\
                                              &                                                                                                                                                                                                                                                      & dropout\_p\_hidden              & Real        & 0 - 0.9             & 10    \\
                                              &                                                                                                                                                                                                                                                      & constrained\_embedding          & Boolean     & True, False         &       \\ \midrule
\multirow{3}{*}{\grureminder}  & \multirow{3}{*}{\begin{tabular}[c]{@{}l@{}}The optimal set of hyperparameters of \gru\\ reminders = True\\ remind\_strategy = hybrid\end{tabular}}                                                                                                                                                                                                                                    & \remindsess      & Integer     & 1 - 10              & 10    \\
                                              &                                                                                                                                                                                                                                                      & \wbase           & Integer     & 1 - 10              & 10    \\
                                              &                                                                                                                                                                                                                                                      & \wirec           & Integer     & 0 - 9               & 10    \\ \midrule
\multirow{3}{*}{\narm}         & \multirow{3}{*}{\begin{tabular}[c]{@{}l@{}}epochs = 20\\ layer\_size = 100\end{tabular}}                                                                                                                                                                                                       & factors                         & Integer     & 50, 100             &       \\
                                              &                                                                                                                                                                                                                                                      & \multirow{2}{*}{learning\_rate} & Real        & 0.01-0.001          & 10    \\
                                              &                                                                                                                                                                                                                                                      &                                 & Real        & 0.001-0.0001        & 10    \\ \midrule
\multirow{3}{*}{\narmreminder} &  \multirow{3}{*}{\begin{tabular}[c]{@{}l@{}}The optimal set of hyperparameters of \narm\\ reminders = True\\ remind\_strategy = hybrid\end{tabular}}                                                                                                                                                                                                                                                    & \remindsess      & Integer     & 1 - 10              & 10    \\
                                              &                                                                                                                                                                                                                                                      & \wbase           & Integer     & 1 - 10              & 10    \\
                                              &                                                                                                                                                                                                                                                      & \wirec           & Integer     & 0 - 9               & 10    \\ \midrule
\multirow{9}{*}{\hgru}         & \multirow{9}{*}{\begin{tabular}[c]{@{}l@{}}session\_layers = 100\\ user\_layers = 100\\ loss = top1\\ (as in the original paper)\end{tabular}}                                                                                                                                                        & user\_propagation\_mode         & Categorical & init, all           &       \\
                                              &                                                                                                                                                                                                                                                      & \multirow{2}{*}{learning\_rate} & Real        & 0.01-0.1            & 10    \\
                                              &                                                                                                                                                                                                                                                      &                                 & Real        & 0.1-0.5             & 5     \\
                                              &                                                                                                                                                                                                                                                      & momentum                        & Real        & 0 - 0.9             & 10    \\
                                              &                                                                                                                                                                                                                                                      & final\_activation               & Categorical & linear, relu, tanh  &       \\
                                              &                                                                                                                                                                                                                                                      & dropout\_p\_hidden\_usr         & Real        & 0 - 0.9             & 10    \\
                                              &                                                                                                                                                                                                                                                      & dropout\_p\_hidden\_ses         & Real        & 0 - 0.9             & 10    \\
                                              &                                                                                                                                                                                                                                                      & dropout\_p\_init                & Real        & 0 - 0.9             & 10    \\
                                              &                                                                                                                                                                                                                                                      & batch\_size                     & Integer     & 50, 100             &       \\ \midrule
\multirow{6}{*}{\iirnn}        & \multirow{6}{*}{\begin{tabular}[c]{@{}l@{}}max\_epoch = 100\\ for \retailrocket, \lastfm and \cosmetics\\  max\_epoch = 20 for \xing\\ because of  the computational complexity \end{tabular}}                                                                                                               & dropout\_pkeep                  & Real        & 0.1 - 1             & 10    \\
                                              &                                                                                                                                                                                                                                                      & \multirow{2}{*}{learning\_rate} & Real        & 0.01 - 0.001        & 10    \\
                                              &                                                                                                                                                                                                                                                      &                                 & Real        & 0.001 - 0.0001      & 10    \\
                                              &                                                                                                                                                                                                                                                      & embedding\_size                 & Integer     & 50, 100             &       \\
                                              &                                                                                                                                                                                                                                                      & max\_session\_representations   & Integer     & 1, 5, 10, 15, 20    &       \\
                                              &                                                                                                                                                                                                                                                      & use\_last\_hidden\_state        & Boolean     & True, False         &       \\ \midrule
\multirow{4}{*}{\shan}         & \multirow{4}{*}{\begin{tabular}[c]{@{}l@{}}global\_dimension = 100\\ epochs = 100 \\ (as in the original paper)\end{tabular}}                                                                                                                                                                      &                                 &             &                     &       \\
                                              &                                                                                                                                                                                                                                                      & lambda\_uv                      & Real        & 0.01, 0.001, 0.0001 &       \\
                                              &                                                                                                                                                                                                                                                      & lambda\_a                       & Integer     & 1, 10, 50           &       \\
                                              &                                                                                                                                                                                                                                                      &                                 &             &                     &       \\ \midrule
\multirow{3}{*}{\ncsf}         & \multirow{3}{*}{}                                                                                                                                                                                                                                    & max\_nb\_his\_sess              & Integer     & 0, 1, 2, 5, 10      &       \\
                                              &                                                                                                                                                                                                                                                      & att\_alpha                      & Real        & 0.01, 0.1, 1, 10    &       \\
                                              &                                                                                                                                                                                                                                                      & window\_sz                      & Integer     & 1 - 10              & 10    \\ \midrule
\multirow{5}{*}{\nsar}         & \multirow{8}{*}{\begin{tabular}[c]{@{}l@{}}epochs = 20\\ keep\_pr = 0.25 \\ (as in the original paper)\\  batch\_size = 64\\ for \retailrocket, \lastfm and \cosmetics \\ (as in the original paper) \\ batch\_size =  32 for \xing\\ because of memory limitations\end{tabular}}  &                                 &             &                     &       \\
                                              &                                                                                                                                                                                                                                                      &                                 &             &                     &       \\
                                              &                                                                                                                                                                                                                                                      & \multirow{2}{*}{learning\_rate} & Real        & 0.001 - 0.01        & 10    \\
                                              &                                                                                                                                                                                                                                                      &                                 & Real        & 0.01 - 0.05         & 5     \\
                                              &                                                                                                                                                                                                                                                      & hidden\_units                   & Integer     & 50, 100             &       \\
                                              &                                                                                                                                                                                                                                                      &                                 &             &                     &       \\
                                              &                                                                                                                                                                                                                                                      &                                 &             &                     &       \\
                                              &                                                                                                                                                                                                                                                      &                                 &             &                     &       \\ \bottomrule

\end{tabular}}
\end{table}

\begin{table}[ht]
	\centering
	\caption{Optimal hyperparameters for simple methods.}
	\label{tab:opt_hyp_simple}
	\Huge
	\def\arraystretch{0.5}
	\resizebox{0.8\textwidth}{!}{%
		\begin{tabular}{llllll}
			\toprule
			Algorithm                                & Hyperparameter          & \retailrocket            & \xing                   & \cosmetics               & \lastfm           \\
			\midrule
			\multirow{2}{*}{\sr}                     & steps                   & 15                       & 25                      & 15                       & 8                 \\
			                                         & weighting               & quadratic                & quadratic               & div                      & quadratic         \\
			\midrule
			\multirow{9}{*}{\sr with extensions}     & \textbf{best extension} & \textbf{\usrreminder}    & \textbf{\usrreminder}   & \textbf{\usrreminder}    & \textbf{\usr}     \\
			                                         & steps                   & 12                       & 30                      & 15                       & 20                 \\
			                                         & weighting               & quadratic                & quadratic               & div                      & quadratic         \\
			                                         & boost\_own\_sessions    & 3.1                      & 1.9                     & 3.7                      & 3.1                 \\
			                                         & reminders               & True                     & True                    & True                     & -            \\
			                                         & remind\_strategy        & hybrid                   & hybrid                  & hybrid                   & -            \\
			                                         & remind\_sessions\_num   & 2                        & 6                       & 9                        & -                 \\
			                                         & weight\_base            & 5                        & 8                       & 8                        & -                 \\
			                                         & weight\_IRec            & 3                        & 4                       & 3                        & -                 \\
			\midrule
			\multirow{5}{*}{\vsknn}                  & k                       & 50                       & 100                     & 100                      & 50                \\
			                                         & sample\_size            & 500                      & 500                     & 10000                    & 500               \\
			                                         & weighting               & log                      & log                     & quadratic                & quadratic         \\
			                                         & weighting\_score        & linear                   & quadratic               & div                      & quadratic         \\
			                                         & idf\_weighting          & 10                       & 10                      & 10                       & 5                 \\
			\midrule
			\multirow{14}{*}{\vsknn with extensions} & \textbf{best extension} & \textbf{\uvsknnreminder} & \textbf{\vsknnreminder} & \textbf{\uvsknnreminder} & \textbf{\uvsknn}  \\
			                                         & k                       & 1500                     & 100                     & 1500                     & 50                \\
			                                         & sample\_size            & 1000                     & 500                     & 10000                    & 500               \\
			                                         & weighting               & log                      & log                     & quadratic                & quadratic         \\
			                                         & weighting\_score        & linear                   & quadratic               & div                      & quadratic         \\
			                                         & idf\_weighting          & 1                        & 10                      & 10                       & 1                 \\
			                                         & extend\_session\_length & 8                        & -                       & 2                        & 3                 \\
			                                         & boost\_own\_sessions    & 0.1                      & -                       & 0.9                      & 2.5               \\
			                                         & reminders               & True                     & True                    & True                     & -            \\
			                                         & remind\_strategy        & hybrid                   & hybrid                  & hybrid                   & -                 \\
			                                         & remind\_sessions\_num   & 4                        & 8                       & 10                       & -                 \\
			                                         & weight\_base            & 8                        & 2                       & 9                        & -                 \\
			                                         & weight\_IRec            & 1                        & 1                       & 2                        & -                 \\
			                                         & weight\_SSim            & 1                        & 0                       & 3                        & -                 \\
			\midrule
			\multirow{5}{*}{\stan}                   & k                       & 1500                     & 100                     & 500                      & 100               \\
			                                         & sample\_size            & 2500                     & 10000                   & 2500                     & 10000             \\
			                                         & lambda\_spw             & 0.905                    & 0.4525                  & 0.905                    & 0.00001           \\
			                                         & lambda\_snh             & 100                      & 80                      & 40                       & 80                \\
			                                         & lambda\_inh             & 0.4525                   & 0.4525                  & 0.4525                   & 3.62              \\
			\midrule
			\multirow{14}{*}{\stan with extensions}  & \textbf{best extension} & \textbf{\staner}         & \textbf{\stanr}         & \textbf{\stanebr}        & \textbf{\stanebr} \\
			                                         & k                       & 200                      & 100                     & 1500                     & 100               \\
			                                         & sample\_size            & 1000                     & 10000                   & 5000                     & 2500              \\
			                                         & lambda\_spw             & 0.905                    & 0.4525                  & 0.905                    & 0.00001           \\
			                                         & lambda\_snh             & 100                      & 80                      & 100                      & 100               \\
			                                         & lambda\_inh             & 0.905                    & 0.4525                  & 7.24                     & 7.24              \\
			                                         & extend\_session\_length & 2                        & -                       & 2                        & 17                \\
			                                         & boost\_own\_sessions    & -                        & -                       & 1.9                      & 2.7               \\
			                                         & reminders               & True                     & True                    & True                     & True            \\
			                                         & remind\_strategy        & hybrid                   & hybrid                  & hybrid                   & hybrid            \\
			                                         & remind\_sessions\_num   & 9                        & 3                       & 4                        & 3                 \\
			                                         & weight\_base            & 10                       & 10                      & 10                       & 5                 \\
			                                         & weight\_IRec            & 3                        & 2                       & 1                        & 0                 \\
			                                         & weight\_SSim            & 2                        & 1                       & 1                        & 6                 \\
			\midrule
			\multirow{8}{*}{\vstan}                  & k                       & 200                      & 1500                    & 500                      & 1000              \\
			                                         & sample\_size            & 5000                     & 10000                   & 1000                     & 5000              \\
			                                         & similarity              & vec                      & cosine                  & cosine                   & cosine            \\
			                                         & lambda\_spw             & 1.81                    & 3.62                    & 3.62                     & 1.81              \\
			                                         & lambda\_snh             & 40                       & 20                      & 80                       & 100               \\
			                                         & lambda\_inh             & 0.905                    & 0.4525                   & 0.4525                  & 1.81              \\
			                                         & lambda\_ipw             & 0.905                    & 0.4525                   & 0.905                    & 0.0001             \\
			                                         & lambda\_idf             & False                    & 10                      & False                    & False             \\
			\midrule
			\multirow{17}{*}{\vstan with extensions} & \textbf{best extension} & \textbf{\vstanebr}       & \textbf{\vstanr}        & \textbf{\vstanebr }      & \textbf{\vstaneb} \\
			                                         & k                       & 2000                     & 1500                    & 500                      & 1000              \\
			                                         & sample\_size            & 10000                    & 10000                   & 1000                     & 10000             \\
			                                         & similarity              & cosine                   & cosine                  & cosine                   & cosine            \\
			                                         & lambda\_spw             & 0.905                    & 3.62                   & 0.905                    & 0.4525            \\
			                                         & lambda\_snh             & 80                       & 20                      & 80                       & 100               \\
			                                         & lambda\_inh             & 1.81                     & 0.4525                  & 0.4525                   & 3.62              \\
			                                         & lambda\_ipw             & 3.62                     & 0.4525                  & 3.62                     & 0.4525            \\
			                                         & lambda\_idf             & 5                        & 10                      & 1                        & 5                 \\
			                                         & extend\_session\_length & 5                        & -                       & 1                        & 7                 \\
			                                         & boost\_own\_sessions    & 0.1                      & -                       & 3.1                      & 3.7               \\
			                                         & reminders               & True                     & True                    & True                     & -            \\
			                                         & remind\_strategy        & hybrid                   & hybrid                  & hybrid                   & -                 \\
			                                         & remind\_sessions\_num   & 2                        & 3                       & 5                        & -                 \\
			                                         & weight\_base            & 6                        & 9                       & 7                        & -                 \\
			                                         & weight\_IRec            & 2                        & 1                       & 1                        & -                 \\
			                                         & weight\_SSim            & 0                        & 5                       & 0                        & -                 \\
			\bottomrule&
			
		\end{tabular}}
\end{table}

\begin{table}[ht]
	\centering
	\caption{Optimal hyperparameters for neural methods.}
	\label{tab:opt_hyp_neural}
	
	\Huge
	\def\arraystretch{0.7}
	\resizebox{0.8\textwidth}{!}{%
		\begin{tabular}{llllll}
			\toprule
			Algorithm                      & Hyperparameter                & \retailrocket & \xing    & \cosmetics & \lastfm \\
			\midrule
			\multirow{6}{*}{\gru}          & learning\_rate                & 0.08          & 0.05     & 0.03       & 0.04    \\
			& momentum                      & 0.1           & 0.6      & 0.3        & 0.1     \\
			& loss                          & top1-max      & top1-max & bpr-max        & bpr-max \\
			& final\_act                    & linear        & elu-0.5  & linear     & linear  \\
			& dropout\_p\_hidden            & 0.7           & 0.8      & 0.7        & 0       \\
			& constrained\_embedding        & True          & True     & True       & False   \\
			\midrule
			\multirow{9}{*}{\grureminder}  & learning\_rate                & 0.08          & 0.05     & 0.03       & 0.04    \\
			& momentum                      & 0.1           & 0.6      & 0.3        & 0.1     \\
			& loss                          & top1-max      & top1-max & bpr-max    & bpr-max \\
			& final\_act                    & linear        & elu-0.5  & linear     & linear  \\
			& dropout\_p\_hidden            & 0.7           & 0.8      & 0.7        & 0       \\
			& constrained\_embedding        & True          & True     & True       & False   \\
			& reminders               & True                     & True                    & True                     & True            \\
			& remind\_strategy              & hybrid        & hybrid   & hybrid     & hybrid  \\
			& remind\_sessions\_num         & 3             & 3        & 2          & 4       \\
			& weight\_base                  & 9             & 9        & 7          & 4       \\
			& weight\_IRec                  & 2             & 4        & 3          & 0       \\
			\midrule
			\multirow{2}{*}{\narm}         & factors                       & 50            & 100      & 100        & 100     \\
			& learning\_rate                & 0.01          & 0.007    & 0.007      & 0.007   \\
			\midrule
			\multirow{5}{*}{\narmreminder} & factors                       & 50            & 100      & 100           & 100     \\
			& learning\_rate                & 0.01          & 0.007    & 0.007           & 0.007   \\
			& reminders               & True                     & True                    & True                     & True            \\
			& remind\_strategy              & hybrid        & hybrid   &  hybrid          & hybrid  \\
			& remind\_sessions\_num         & 4             & 7        &  3          & 6       \\
			& weight\_base                  & 10            & 7        &  7          & 3       \\
			& weight\_IRec                  & 7             & 3        &  6          & 0       \\
			\midrule
			\multirow{8}{*}{\hgru}         & user\_propagation\_mode       & all           & all      & init           & all     \\
			& learning\_rate                & 0.06          & 0.08     & 0.04           & 0.09    \\
			& momentum                      & 0.3           & 0.6      & 0.5           & 0.5     \\
			& final\_act                    & linear        & tanh     & linear           & linear  \\
			& dropout\_p\_hidden\_usr       & 0.4           & 0.8      & 0.5           & 0.7     \\
			& dropout\_p\_hidden\_ses       & 0.3           & 0        & 0.3           & 0.1     \\
			& dropout\_p\_init              & 0.4           & 0.6      & 0.1           & 0.3     \\
			& batch\_size                   & 50            & 100      & 50           & 50      \\
			\midrule
			\multirow{5}{*}{\iirnn}        & dropout\_pkeep                & 0.4           & 0.6      &  0.5          & 0.6     \\
			& learning\_rate                & 0.002         & 0.002    &  0.001          & 0.001   \\
			& embedding\_size               & 100           & 100      & 100           & 100     \\
			& max\_session\_representations & 15            & 1        &  1          & 20      \\
			& use\_last\_hidden\_state      & False         & False    &  True          & True    \\
			\midrule
			\multirow{2}{*}{\shan}         & lambda\_uv                    & 0.01          & 0.01     &  0.01          & 0.01    \\
			& lambda\_a                     & 1             & 1        &  1          & 10      \\
			\midrule
			\multirow{3}{*}{\ncsf}         & max\_nb\_his\_sess            & 5             & 0        &  0          & 0       \\
			& att\_alpha                    & 10            & 10       &  1          & 1       \\
			& window\_sz                    & 2             & 2        &   3         & 1       \\
			\midrule
			\multirow{2}{*}{\nsar}         & learning\_rate                & 0.01          & 0.004    &  0.007          & 0.003   \\
			& hidden\_units                 & 100           & 100      &  100          & 100     \\
			\bottomrule&
	\end{tabular}}
\end{table}

\begin{table}[t]
	\centering
	\caption[Results of the performance comparison]{\textcolor{blue}{Running times
	on the \retailrocket dataset.}}
	\label{tab:times_retail}
	\resizebox{0.6\textwidth}{!}{%
		\begin{tabular}{lrr}
			\toprule
			& Training Time (s) & Prediction Time (ms) \\ \midrule
			\cellcolor{1}\textcolor{brown}{\sr}             & \textcolor{brown}{0.42}              & \textcolor{brown}{3.24}              \\
			\cellcolor{1}\textcolor{brown}{\vsknn}          & \textcolor{brown}{0.09}              & \textcolor{brown}{4.42}              \\
			\cellcolor{1}\textcolor{brown}{\stan}           & \textcolor{brown}{0.08}              & \textcolor{brown}{4.29}              \\
			\cellcolor{1}\textcolor{brown}{\vstan}          & \textcolor{brown}{0.09}              & \textcolor{brown}{4.55}              \\
			\cellcolor{1}\usrreminder    & 0.78              & 12.99             \\
			\cellcolor{1}\uvsknnreminder & 0.24              & 20.85             \\
			\cellcolor{1}\staner         & 0.19              & 24.42             \\
			\cellcolor{1}\vstanebr       & 0.24              & 14.99             \\
			\cellcolor{2}\textcolor{brown}{\gru}            & \textcolor{brown}{12.95}             & \textcolor{brown}{4.00}              \\
			\cellcolor{2}\textcolor{brown}{\narm}           & \textcolor{brown}{102.73}            & \textcolor{brown}{5.50}              \\
			\cellcolor{2}\grureminder    & 10.46             & 13.61             \\
			\cellcolor{2}\narmreminder   & 93.90             & 15.07             \\
			\cellcolor{3}\hgru           & 23.92             & 4.07              \\
			\cellcolor{3}\iirnn          & 1307.41           & 87.34             \\
			\cellcolor{3}\ncsf           & 82.54             & 25.81             \\
			\cellcolor{3}\nsar           & 475.46            & 21.51             \\
			\cellcolor{3}\shan           & 1661.60           & 12.32             \\ \bottomrule
	\end{tabular}}
\end{table}

\begin{table}[t]
	\centering
	\caption[Results of the performance comparison]{\textcolor{blue}{Running times
on the \cosmetics dataset.}}
	\label{tab:times_cosmetics}
	\resizebox{0.6\textwidth}{!}{%
		\begin{tabular}{lrr}
			\toprule
			& Training Time (s) & Prediction Time (ms) \\ \midrule
\cellcolor{1}\textcolor{brown}{\sr}             & \textcolor{brown}{0.62}              & \textcolor{brown}{2.64}              \\
\cellcolor{1}\textcolor{brown}{\vsknn}          & \textcolor{brown}{0.14}              & \textcolor{brown}{4.45}              \\
\cellcolor{1}\textcolor{brown}{\stan}           & \textcolor{brown}{0.13}              & \textcolor{brown}{6.06}              \\
\cellcolor{1}\textcolor{brown}{\vstan}          & \textcolor{brown}{0.14}              & \textcolor{brown}{6.81}              \\
\cellcolor{1}\usrreminder    & 1.55              & 8.14              \\
\cellcolor{1}\uvsknnreminder & 0.33              & 14.92             \\
\cellcolor{1}\stanebr        & 0.32              & 18.37             \\
\cellcolor{1}\vstanebr       & 0.33              & 12.42             \\
\cellcolor{2}\textcolor{brown}{\gru}            & \textcolor{brown}{13.56}             & \textcolor{brown}{3.86}              \\
\cellcolor{2}\textcolor{brown}{\narm}           & \textcolor{brown}{159.79}            & \textcolor{brown}{5.62}              \\
\cellcolor{2}\grureminder    & 11.74             & 9.32              \\
\cellcolor{2}\narmreminder   & 135.13            & 21.91             \\
\cellcolor{3}\hgru           & 36.74             & 3.85              \\
\cellcolor{3}\iirnn          & 1034.02           & 59.08             \\
\cellcolor{3}\ncsf           & 40.59             & 10.66             \\
\cellcolor{3}\nsar           & 387.43            & 11.26             \\
\cellcolor{3}\shan           & 2344.50           & 11.58             \\ \bottomrule
	\end{tabular}}
\end{table}

\end{document}